\documentclass[%
reprint,
superscriptaddress,
 amsmath,amssymb,
 aps,
prb,
]{revtex4-2}

\usepackage{graphicx}% Include figure files
\usepackage{dcolumn}% Align table columns on decimal point
\usepackage{bm}% bold math
\usepackage[colorlinks,citecolor=blue]{hyperref} \usepackage{cleveref}% add hypertext capabilities
\hypersetup{colorlinks=true, linkcolor=blue, filecolor=magenta, urlcolor=blue,}
\usepackage[mathlines]{lineno}% Enable numbering of text and display math
\usepackage{comment}
\usepackage{subfigure}
\usepackage{xcolor}
\usepackage{textcomp}

\begin{document}

\preprint{APS/123-QED}

\title{Minimal loop currents in doped Mott insulators}% Force line breaks with \\

\author{Can Cui}\affiliation
{
Institute for Advanced Study, Tsinghua University, Beijing 100084, China }
\author{Jing-Yu Zhao}\affiliation
{
Department of Physics and Astronomy, Johns Hopkins University, Baltimore, Maryland 21218, USA
}
\author{Zheng-Yu Weng}%
 %\email{weng@tsinghua.edu.cn}
\affiliation
{
Institute for Advanced Study, Tsinghua University, Beijing 100084, China }%

\date{\today}% It is always \today, today,

\begin{abstract}  

{For the $t$-$J$ model, variational wave functions can generally be constructed based on an accurate description of antiferromagnetism (AFM) at half‑filling and an exact phase‑string sign structure under doping. The single‑hole‑doped and two‑hole‑doped states, as determined by variational Monte Carlo (VMC) simulations, display sharply contrasting behaviors. The single‑hole state constitutes a ``cat state'' that resonates strongly between a quasiparticle component and a local loop‑current component, with approximately equal weights. In the ground state, the quasiparticle spectral weight $Z_{\mathbf{k}}$ peaks at momenta $\mathbf{k}_0 \equiv (\pm\frac{\pi}{2},\pm\frac{\pi}{2})$. The total‑energy dispersion versus $\mathbf{k}$ agrees remarkably well with the Green function Monte Carlo results throughout the Brillouin zone. 
However, Landau’s one‑to‑one correspondence hypothesis for quasiparticles breaks down here with the incoherent component exhibiting intrinsic magnetization originating from a minimal $2\times2$ loop current that forms a $4\times4$ pattern on the square lattice—a finding in excellent agreement with density matrix renormalization group (DMRG) calculations. In the two‑hole ground state, a new pairing mechanism is revealed: the two holes are automatically fused into a tightly bound object consisting of an incoherent $d_{xy}$ pairing along the diagonal direction by compensating the local loop currents. This hole pair is again a  ``cat state'' that resonates strongly between the incoherent $d_{xy}$ and a coherent $d_{x^2-y^2}$ Cooper channel to gain substantial hopping energy. Its size extends over an area of about $4\times 4$ lattice spacings, much smaller than the divergent AFM correlation length, implying that it should survive as a minimal superconducting building block even in the dilute doping regime. Experimental implications and the generalization to the finite-doping case are briefly addressed.

}    

\end{abstract}

\maketitle

%\tableofcontents

\section{Introduction}

The microscopic mechanism of high-temperature superconductivity in the cuprates has remained controversial since its discovery in 1986 \cite{Bednorz1986}. Although doped Mott insulators are widely believed to capture the essential physics of these materials \cite{Anderson1987,RevModPhys.78.17}, their strong correlations are inherently nonperturbative, leaving the theoretical understanding of their ground-state properties incomplete. Two of the most fundamental open questions are whether or how single-particle excitations violate Landau quasiparticle criteria, and how two doped holes can become intrinsically paired in the doped Mott insulator without invoking an external force like the electron-phonon interaction.

The dilute doping regime is characterized by a long antiferromagnetic (AFM) correlation length, which naturally distinguishes between long‑wavelength renormalization effects and short‑range spin–charge entanglement. In the extreme limit of one or two holes, the AFM background remains thermodynamically unchanged, allowing the local mutual entanglement between the spin environment and the dopants to be cleanly isolated. As this work will demonstrate, the resulting renormalization is strongly non‑perturbative—so much so that it endows a single hole with singular behavior—and, in turn, provides the very mechanism that binds two holes into a compact pair on a scale far smaller than the AFM correlation length. Such a tightly bound pair may thus be regarded as a minimal building block, opening a route to study its interplay with long‑range AFM correlations at finite doping.

Traditionally, analogous to a quasielectron dressed by electron-hole pairs in a normal metal, early studies often described a doped hole as a coherent, Landau-like quasiparticle dressed by a distortion of the spin background—known as the “longitudinal spin polaron effect'' \cite{PhysRevB.2.1324,PhysRevLett.60.2793,PhysRevB.39.6880}. In the self-consistent Born approximation (SCBA), scattering between holes and spin magnons renormalizes the hole's effective mass while its spectral weight remains finite. The calculated energy minimum of the single-hole-doped $t$-$J$ model is located at $\mathbf{k}_0 = (\pm \frac{\pi}{2}, \pm \frac{\pi}{2})$, consistent with exact diagonalization (ED) results on small systems \cite{RevModPhys.66.763,PhysRevB.52.R15711}.

However, recent density matrix renormalization group (DMRG) studies of the single-hole-doped $t$-$J$ model in two dimensions (2D) with open boundary conditions (OBC) and preserved $C_4$ rotational symmetry reveal a twofold ground-state degeneracy characterized by a nontrivial quantum number $L_z = \pm 1$, corresponding to the chirality of “hidden'' charge and spin loop currents in the system \cite{PhysRevB.98.165102}. This short-range loop-current pattern is absent in the semiclassical SCBA approach, indicating a strong correlation effect emerging at the lattice scale. It implies that the single-hole state cannot be simply specified by a momentum $\mathbf{k}$ as a Landau quasiparticle. Similar dual behavior of a single hole in the infrared and ultraviolet limits has also been found in the two-leg ladder $t$-$J$ model \cite{PhysRevB.92.235156,PhysRevB.98.035129},
where a Landau quasiparticle state is only recovered as the spin-spin correlation length becomes sufficiently short along the chain direction in the strongly anisotropic (strong rung) limit.

On general grounds, Anderson argued \cite{Anderson1990,PhysRevLett.18.1049} that when an additional hole is injected into a (doped) Mott insulator, the many-body system can acquire a nonlocal “unrenormalizable Fermi‑surface phase shift'' to accommodate it, leading to an “orthogonality catastrophe'' such that the doped state is no longer Landau-Fermi-liquid-like. Such a nontrivial many‑body phase shift has been explicitly identified as the phase‑string effect in the $t$‑$J$ model \cite{PhysRevLett.77.5102,Weng1997,PhysRevB.77.155102} and the Hubbard model \cite{PhysRevB.90.165120}. It originates from the irreparable string‑like spin mismatch created as the doped hole moves. The spin‑polaron picture mentioned earlier shows \cite{PhysRevB.39.6880} that longitudinal $S^z$ spin mismatches may be repaired through spin‑flip processes, but it overlooks the fact that transverse $S^\pm$ mismatches cannot be simultaneously repaired. The result is a trail of signs $\pm $, depending on whether the hole exchanges places with an up or down spin along its trajectory; this is called the ``phase string'', representing the transverse defect left by hole hopping. Thus, the phase‑string sign structure can be regarded as a generalized many‑body Berry phase accumulated when the hole completes a closed loop, and it plays a key role in the quantum interference among different Feynman paths of the doped holes, fundamentally shaping the many-body hole-spin states of a doped Mott insulator.

\begin{figure}[t]
\includegraphics[width=0.35\textwidth]{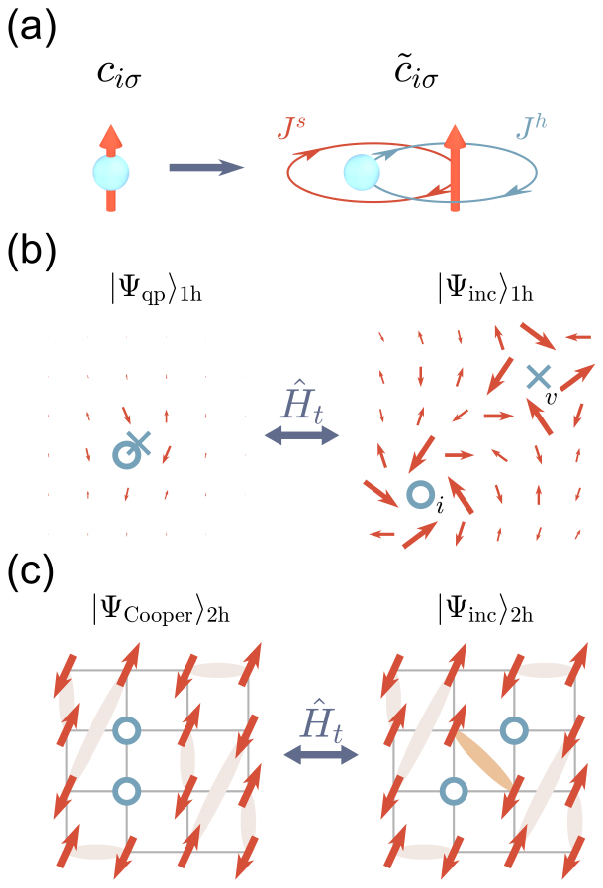}
\caption{\label{fig:string}  
Schematic illustration of doped-hole wavefunctions.
(a) A bare hole of $S^z=\pm 1/2$ is transformed into a twisted hole [Eq. (\ref{eq:twistedc})], a composite of charge and spin that encircle one another in the transverse ($x$-$y$) plane with chirality and orbital angular momentum $L_z=\pm 1$. While the bare hole's motion is blocked by the phase‑string effect, the twisted hole is mobile on the AFM background via the hopping term \cite{PhysRevB.99.205128}. 
(b) The single‑hole variational wave function [Eq.~(\ref{eq:singlehole})] describes a bound state between a twisted hole and an antimeron (centered at the cross). Both exhibit vortex‑like distortions in the spin $x$–$y$ plane with opposite chiralities, thereby removing the logarithmically divergent superexchange energy associated with the twisted hole alone. The bare‑hole state $|\Psi_{\mathrm{qp}}\rangle_{\mathrm{1h}}$ can be recovered when the vortex–antivortex pair annihilates at short distance. This single‑hole state thus forms a quantum ``cat state'', arising from a strong resonance between the quasiparticle component $|\Psi_{\mathrm{qp}}\rangle_{\mathrm{1h}}$ and incoherent component $|\Psi_{\mathrm{inc}}\rangle_{\mathrm{1h}}$ driven by the hopping term.
(c) The two‑hole ground state [Eq.~(\ref{eq:twohole})]. Here, the hopping term again drives a strong resonance between a coherent Cooper‑pair component $|\Psi_{\mathrm{Cooper}}\rangle_{\mathrm{2h}}$ with $d_{x^2-y^2}$ pairing symmetry on the AFM background $|\phi_0\rangle$—a condensate of spin‑singlet pairs on opposite sublattices \cite{Liang1988}—and an incoherent component $|\Psi_{\mathrm{inc}}\rangle_{\mathrm{2h}}$ with 
$d_{xy}$ pairing symmetry, accompanied by the same sublattice spin‑singlet pair (connected by the orange bond), whose amplitude is no longer sign-definite and depends on the hopping history of the holes. } 
\end{figure}

To incorporate this phase-string sign structure in the single-hole problem, the following variational wave function was proposed in previous variational Monte Carlo (VMC) studies \cite{PhysRevB.99.205128}:
\begin{equation}
|\Psi_{\mathrm{G}}\rangle_{1\mathrm{h}}=\sum_{i}\varphi_{\mathrm{h}}(i)\tilde{c}_{i\sigma}|\phi_{0}\rangle, \label{eq:singleholeold}
\end{equation}
where $\tilde{c}_{i\sigma}$ creates a twisted hole
\begin{equation}\label{eq:twistedc}
\tilde{c}_{i\sigma}\equiv c_{i\sigma}e^{\mp i\hat{\Omega}_{i}}
\end{equation} 
on the ``vacuum state'' $|\phi_0\rangle$—the half-filling ground state of the Heisenberg Hamiltonian. 
The phase-string effect is encoded by the phase-shift operator $\hat{\Omega}_i$ in Eq.~(\ref{eq:twistedc}), which generates a transverse spin distortion [cf. Eq.~(\ref{eq:Omega})] or a “transverse spin-polaron'' in the spin background that effectively facilitates hole hopping. This effect manifests as transverse spin and charge currents mutually encircling with chirality [schematically illustrated in Fig.~\ref{fig:string}(a)], resulting in a nontrivial angular momentum $L_z = \pm 1$ \cite{PhysRevB.99.205128} in the ground state, which is consistent with ED and DMRG results \cite{PhysRevB.98.165102}. The calculated quasiparticle spectral weight peaks at $\mathbf{k}_0 = (\pm \frac{\pi}{2}, \pm \frac{\pi}{2})$, but vanishes in the thermodynamic limit. In sharp contrast to the strong suppression of bare hole propagation, however, the “twisted'' quasiparticle $\tilde{c}_{i\sigma}$ can still propagate coherently on top of $|\phi_0\rangle$ with an almost plane-wave-like $\varphi_{\mathrm{h}}(i)$ determined by VMC \cite{PhysRevB.99.205128}. Similarly, the twisted hole state in terms of $\tilde{c}|\phi_0\rangle$ is shown \cite{wang2015variationalwavefunctionanisotropic,PhysRevB.107.085112} to resemble a Bloch-wave propagation in the two-leg ladder system, which is in excellent agreement with DMRG.  

Although the twisted quasiparticle $\tilde{c}_{i\sigma}$ can hop coherently by effectively erasing the singular phase-string effect, the spin-current vortex induced by $e^{\mp i\hat{\Omega}_i}$ can also lead to a logarithmically divergent superexchange energy in the large-sample limit in 2D. To remedy this, one may note that the “twisted hole'' can further induce an “antimeron'' excitation \cite{PhysRevLett.90.157003,PhysRevB.67.115103} in the spin background, creating a spin current of opposite chirality centered at a site $v$ such that
\begin{equation}
|\phi_0\rangle \rightarrow |\tilde{\phi}_0\rangle = e^{\pm i\hat{\Omega}_v}|\phi_0\rangle~.
\end{equation}
Namely, it would be self-consistent to improve the single-hole wave function Ansatz in Eq.~(\ref{eq:singleholeold}) by replacing $|\phi_0\rangle$ with $|\tilde{\phi}_0\rangle$  to account for the feedback effect of the twisted hole’s motion on the spin background as schematically illustrated in Fig.~\ref{fig:string}(b). 

In this paper, we systematically examine such a new Ansatz state using VMC and compare it with numerical results. 
While the ground state shows the same quantum number $L_z=\pm 1$, the variational ground-state energy is significantly improved in this new single-hole wave function. The quasiparticle component has a spectral weight peaking at the four $\mathbf{k}_0$ points in the ground state, and its excitation at $\mathbf{k}$ exhibits an energy dispersion in excellent agreement with large-scale numerical simulations of the $t$-$J$ model across the entire Brillouin zone. 
In particular, it has been shown to be a ``cat state'' strongly resonating between a quasiparticle component and an incoherent component with roughly equal weights [cf. Fig.~\ref{fig:string}(b)]. Here, the charge and spin loop currents discovered by DMRG \cite{PhysRevB.98.165102} are consistently reproduced by the incoherent component of the wave function, where minimal $2\times2$ loop charge currents form a $4\times4$ pattern on the square lattice with magnetization $\sim 0.1\mu_B$. 
In other words, the single hole is no longer a Bloch wave of a ``point-like'' particle dressed by the many-body spin-polaron effect. Instead, the doped hole behaves as a resonating hopping process, between the point-like quasiparticle and a ``closed string'' on a minimal $2 \times 2$ plaquette on the square lattice.   

This dual structure of the single-hole ``cat state'' illustrates the non-Landau-quasiparticle nature of the wave function: since the quasiparticle component alone cannot specify the basic properties of the wave function—which also includes the incoherent component—Landau’s one-to-one correspondence principle is clearly violated here. Experimentally, its implications are significant: although single-electron detectors like angle-resolved photoemission spectroscopy (ARPES) and scanning tunneling spectroscopy (STS) can directly probe the quasiparticle component, the loop-current component is like “dark matter'' to ARPES/STS measurements; its crucial presence in the single-hole state is only revealed via numerical DMRG. Thus, the “dispersion'' of the quasiparticle observed in ARPES can be mistakenly attributed to a single-particle property of a band structure with conventional many-body renormalization, while in reality it is driven by the loop-current component via the hopping term. In contrast, if one artificially switches off the phase-string sign structure in the $t$-$J$ model [cf. the $\sigma\cdot t$-$J$ model discussed in Appendix \ref{app:longitudinal}], the incoherent component can also be turned off by setting $\hat{\Omega}_i=0$ in Eq.~(\ref{eq:singleholeold}), yielding a correct and excellent VMC result in terms of the quasiparticle component alone. 

A two-hole variational ground state can be naturally constructed by placing the second hole at the antimeron position in the modified single-hole wave function \cite{PhysRevX.12.011062}. Such a two-hole state [cf. Eq.~(\ref{eq:twohole})] shows strong binding of the two holes with $d$-wave symmetry. The charge/spin loop currents present in the single-hole case are completely compensated in the non-degenerate two-hole ground state with $L_z= 2$ (mod 4) under $C_4$ symmetry. In fact, it is still a ``cat state'' composed of a Cooper pair with $d_{x^2-y^2}$ symmetry and an incoherent component with $d_{xy}$ pairing symmetry, accompanied by an additional spin-singlet excitation [cf. Fig.~\ref{fig:string}(c)]. Again, a predominant hopping energy is gained through strong resonance between the two components, meaning that such a two-hole “fusion'' represents a new pairing mechanism that cannot be simplified as a Cooper pairing instability of two quasiparticles in a perturbative manner. Unlike the single-hole case, we show that although the incoherent component is absent in the low-energy branch of STS, it can still appear in the high-energy branch of the single-particle spectral function (in an inverse ARPES or STS on the positive bias side) above the pairing energy scale. Such a hole pair approximately occupies an area of about $4 a_0 \times 4 a_0$ ($a_0$ is the lattice constant) to constitute a minimal superconducting (SC) building block, consistent with recent STS experiments \cite{Ye2023,Li2023,ye2023visualizingzhangricesingletmolecular}.

The remainder of this paper is structured as follows. In Sec.~\ref{sec:2}, we introduce the $t$-$J$ model and present the optimized single-hole variational Ansatz based on the phase-string sign structure. Using this Ansatz, we compute the single-hole dispersion and show that it successfully captures the quasiparticle degrees of freedom. Benchmarking against DMRG results, we further calculate both spin and hole current patterns, revealing a $2\times2$ vortex structure in the charge current that produces a local magnetic moment, as quantified numerically. In Sec.~\ref{sec:3}, we analyze the structure of the single-hole wave function, clarify the origin of the charge and spin currents, and demonstrate the resonant coupling between the quasiparticle and incoherent components. Sec.~\ref{sec:4} examines the two-hole ground state, where we distinguish its two constituent components using hole-hole correlators, pair-order parameters, and single-particle spectral functions. Finally, Sec.~\ref{sec:6} provides a summary and concluding remarks. 

\section{Single-hole ground state: VMC vs. DMRG}\label{sec:2}

\subsection{The $t$-$J$ model}

One of the simplest models of doped Mott insulators is the $t$-$J$ model, whose Hamiltonian is given by $H_{t\text{-}J}=\mathcal{P}_s (H_t + H_J) \mathcal{P}_s$. The expressions for the hopping term $H_t$ and superexchange term $H_J$ are listed below:
\begin{eqnarray}
H_{t}=-t\sum_{\langle ij\rangle,\sigma}(c_{i\sigma}^{\dagger}c_{j\sigma}+\mathrm{H.c.})~,\label{eq:Ht}
\\
H_J=J\sum_{\langle ij\rangle}\biggl(\mathbf{S}_i\cdot\mathbf{S}_j-\frac{1}{4}n_in_j\biggr)~.\label{eq:HJ}
\end{eqnarray}
Here $\mathcal{P}_s$ is the projection operator imposing the no-double-occupancy constraint on the hole-doped side: $\sum_{\sigma}c^{\dagger}_{i\sigma}c_{i\sigma}\leq 1$ at each site, leading to the strong correlation effect. The superexchange term $J=1$ is set as the energy unit, and the hopping integral $t = 3$ is fixed throughout the paper unless explicitly specified. The size of the square lattice is $N_x \times N_y$ under the OBC.

At half-filling, $H_{t\text{-}J}$ reduces to $H_J$ as the Heisenberg model, whose ground state $|\phi_0\rangle$ can be highly accurately simulated by the Liang-Doucot-Anderson variational wave function \cite{Liang1988} using the Monte Carlo method (cf. Ref. \cite{PhysRevB.82.024407}). Here $|\phi_0\rangle$ is composed of the condensate of spin singlets or resonating-valence-bond (RVB) pairs that only connect opposite sublattices. It is spin-singlet and translationally invariant for a finite-size square lattice, whose spin-spin correlation length diverges in two dimensions (2D) in the thermodynamic limit \cite{Liang1988}. 

\subsection{Single-hole wave function \emph{Ansatz}}

\begin{figure*}[t]
\includegraphics[width=1.0\textwidth]{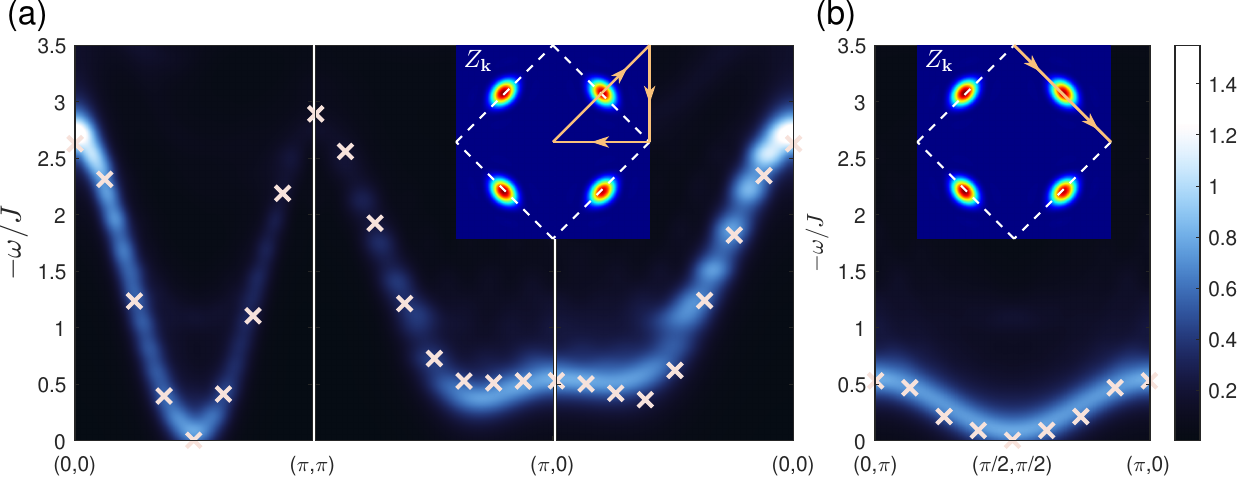}% Here is how to import EPS art
\caption{\label{fig:Akwneg} 
Single-particle spectral function $A^n(\mathbf{k},\omega)$ defined in Eq.~(\ref{eq:negbias}) (with $\eta=0.15$). Intensity of the spectral function plotted along high-symmetry momentum directions in the Brillouin zone (scan paths indicated by yellow arrows in the inset) for energies $\omega<0$. Results are from VMC calculations on a $14\times 14$ system with $t/J = 2.5$, where the energy bottom is set to $\omega=0$. Crosses denote benchmark dispersion data from Green Function Monte Carlo simulations \cite{BONINSEGNI1994330}. Inset: Spectral weight $Z_{\mathbf{k}}$ for the ground state, showing pronounced peaks at momenta $\mathbf{k}_0$. The white dashed lines indicate the magnetic Brillouin zone boundaries.}
\end{figure*}

The simplest nontrivial problem of the $t$-$J$ model concerns the single-hole-doped ground state and low-lying excitations. As outlined in the Introduction, we will investigate the following single-hole wave function Ansatz:
\begin{equation}
|\Psi \rangle_{1\mathrm{h}}=\sum_{i,v,m=\pm 1}\varphi_{m}(i,v)c_{i\sigma}e^{-im\left(\hat{\Omega}_{i}-\hat{\Omega}_{v}\right)}|\phi_{0}\rangle, \label{eq:singlehole}
\end{equation}
in which a spin-$\sigma/2$ ($\sigma=\pm 1$) electron is removed from the 2D spin background $|\phi_0\rangle$. 

In Eq.~(\ref{eq:singlehole}), $\hat\Omega_{i}$ represents the many-body phase shift induced by the doped hole, defined as \cite{PhysRevB.99.205128}:
\begin{equation}
\hat{\Omega}_i=\sum_{l(\neq i)}\theta_i(l)n_{l\downarrow},\label{eq:Omega}
\end{equation}
where $n_{l\downarrow}$ is the number of $\downarrow$-spins on site $l$, and $\theta_i(l)= \mathrm{Im}\ln(z_{l}-z_{i})$ is the statistical angle between the horizontal axis and the vector from the hole to a down spin at $l$ (denoted by the complex coordinate $z_l = x_l + i y_l$ in 2D).
By modifying the bare quasiparticle $c_{i\sigma}$ to the twisted quasiparticle $\tilde{c}_{i\sigma}$ defined in Eq.~(\ref{eq:twistedc}), each down-spin in $|\phi_0\rangle$ is rotated by an angle $\theta_i(l)$ around the $\hat{z}$ quantization axis. In effect, a $2\pi$-vortex centered at the hole site $i$ is formed by the down spins. It has been shown previously \cite{PhysRevB.99.205128} that such a vortex-like twist of spins in the $xy$-plane can effectively compensate the quantum frustration caused by the singular phase-string effect, thereby facilitating hole hopping for the single-hole state in Eq.~(\ref{eq:singleholeold}).

However, the twisted hole $\tilde{c}_{i\sigma}$ alone also generates a nondissipative neutral spin current around it in the AFM long-range-ordered spin background $|\phi_0\rangle$. Generally, such a superfluid-like vortex will induce a logarithmically divergent self-energy due to the spin-vortex configuration in the large sample-size limit. This effect was not fully addressed in previous finite-size calculations \cite{PhysRevB.99.205128}, which primarily focused on the hopping process. As noted before, an ``antimeron" spin configuration \cite{PhysRevLett.90.157003,PhysRevB.67.115103} with opposite chirality should be spontaneously generated in the spin background. 
This is now incorporated into the wave function Ansatz in Eq.~(\ref{eq:singlehole}) through the factor $e^{im\hat\Omega_v}$. Here, $v$ denotes the location of the antimeron center, which in our variational study will be placed at the center of each plaquette on the square lattice.

Finally, the variational parameters $\varphi_{m}(i,v)$ in Eq.~(\ref{eq:singlehole}) are determined by diagonalizing the effective matrices of $H_{t\text{-}J}$ under the basis set defined in Eq.~(\ref{eq:singlehole}), following the method in Ref. \cite{PhysRevB.110.155111,PhysRevB.111.104502} . Here the amplitude $|\varphi_{m}(i,v)|^2$ governs the distance between the twisted hole and the antimeron. Based on \emph{Ansatz} Eq.~(\ref{eq:singlehole}), the properties of the single-hole ground state and excited states can be systematically investigated in the following, using the unbiased DMRG method as the benchmark (cf. Appendix ~\ref{app:secA}). 
For the $8 \times 8$ system, the variational ground energy is found to be lowered by $1.675$ as compared to the previous variational wave function without antimeron \cite{PhysRevB.99.205128}, with the total energy $E_{\mathrm{tot}}(\mathrm{VMC})=-71.467$ close to the DMRG result $E_{\mathrm{tot}}(\mathrm{DMRG})=-72.979$. It is noted that the total energy may be further improved with incorporating the aforementioned longitudinal spin-polaron effect (cf. Appendix ~\ref{app:longitudinal}), which nevertheless does not change the quantum number (angular momentum) $L_z=\pm 1$ and the degeneracies of the single-hole ground state. Thus, our following analyses will mainly focus on the simpler wave function in Eq.~(\ref{eq:singlehole}).

\subsection{Quasiparticle spectral weight and dispersion}

\begin{figure*}[t]
\includegraphics[width=0.6\textwidth]{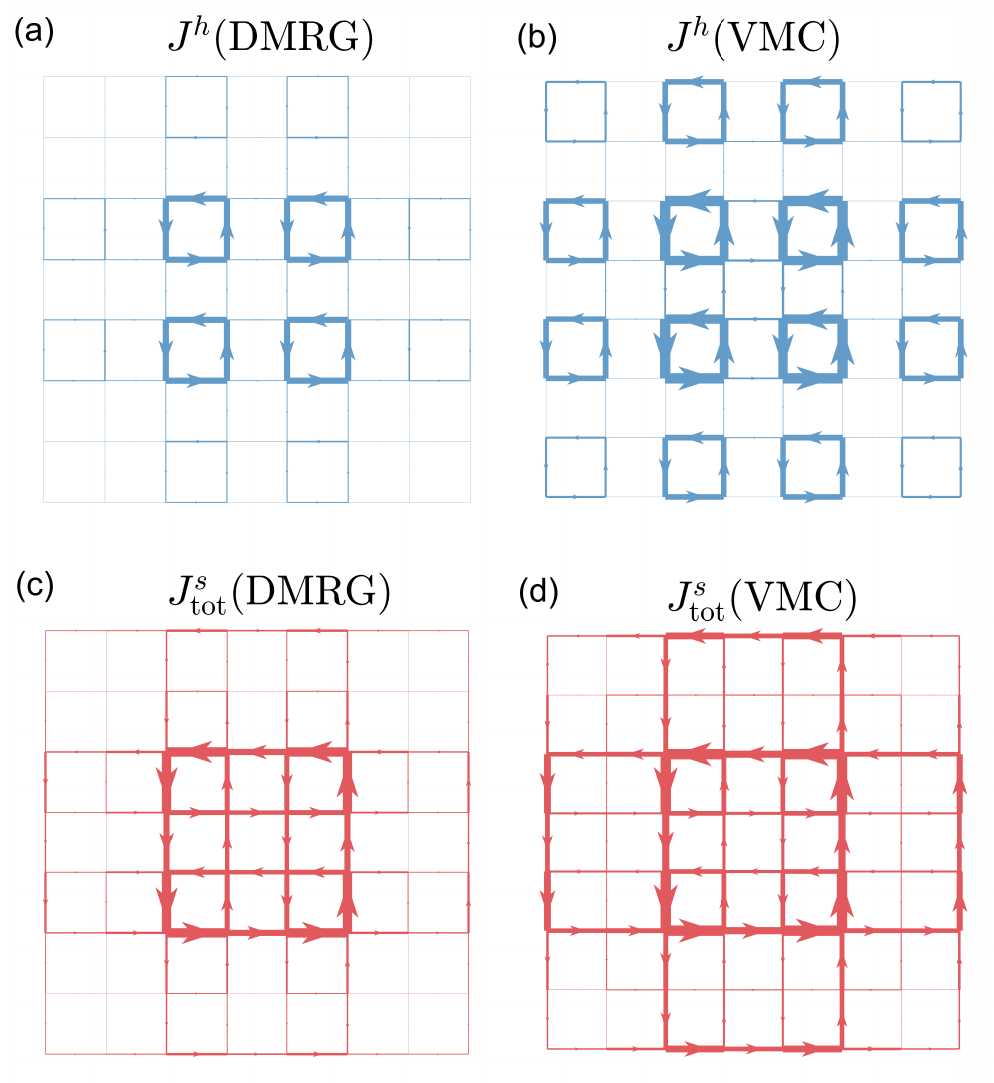}% Here is how to import EPS art
\caption{\label{fig:currenttotal} 
Hole and spin currents for the degenerate one-hole ground state ($L_z = 1$ and $S^z =-\frac{1}{2}$). Calculated hole current $J^h$ (a, b) and total spin current $J^s_{\mathrm{tot}}$ (c, d) on an $8\times 8$ lattice. Results from DMRG (a, c) and VMC (b, d) are shown, respectively. Arrow thickness indicates relative current strength; hole and spin currents are displayed on separate scales.} 
\end{figure*}

The quasiparticle spectral weight
\begin{equation}
Z_{\mathbf{k}}=\left|\langle\phi_0|c^{\dagger}_{\mathbf{k}\sigma}|\Psi \rangle_{1\mathrm{h}}\right|^2~
\label{eq:specweight_def}
\end{equation}
calculated by VMC shows four peaks at momenta $\mathbf{k}_0= (\pm\frac{\pi}{2},\pm\frac{\pi}{2})$ in the ground state $|\Psi_{\mathrm{G}}\rangle_{1\mathrm{h}}$ (see the inset of Fig.~\ref{fig:Akwneg}), which is consistent with ED \cite{PhysRevB.99.205128} and DMRG results \cite{PhysRevB.98.165102}. The single-hole ground state Eq. (\ref{eq:singlehole}) is degenerate, characterized by the angular momentum $L_z=\pm 1$ and spin $S^z=\pm 1/2$.  Here, the total angular momentum $L_z=\pm 1$ means that it is a superposition at four $\mathbf{k}_0$ that is present in the ground state of Eq.~(\ref{eq:singlehole}) as the quasiparticle component.  

The dispersion of the single hole is characterized by the single-particle spectral function defined as follows:
\begin{equation}
A^{n}(\mathbf{k},\omega)=-\mathrm{Im}\sum_{n}\frac{\left| _{\mathrm{1h}}\langle\Psi(n)|c_{\mathbf{k}\sigma}|\phi_{0}\rangle\right|^{2}}{\omega+[E_{\mathrm{1h}}(n)-E_{\mathrm{0}}+\mu_n]+i\eta}~.
\label{eq:negbias}
\end{equation}
Here, $|\Psi(n)\rangle_{\mathrm{1h}}$ denotes the $n$-th excited single-hole state with eigenenergy $E_{\mathrm{1h}}(n)$, $E_{\mathrm{0}}$ is the energy of the half-filling ground state $|\phi_0\rangle$ and $\eta $ is a broadening parameter.  
The chemical potential $\mu_n$ is set to align the single-hole ground-state energy with the zero of energy such that the spectral function $A^n(\mathbf{k},\omega)$ has non-vanishing weight at negative energies.

Figure \ref{fig:Akwneg} displays the spectral function $A^n(\mathbf{k},\omega)$ along high-symmetry lines of the Brillouin zone. The peak at $Z_{\mathbf{k}}$ shows a quasiparticle ``dispersion'' in agreement with the Green function Monte Carlo calculations (beige cross) \cite{BONINSEGNI1994330} (note that $t/J=2.5$ is used here in order to make a direct comparison) throughout the Brillouin zone along the scans indicated in the insets: (1) the dispersion minima lie at the four momenta $\mathbf{k}_0$; (2) the velocity along the nodal direction is much larger than that along the antinodal direction; (3) a relative flat segment is present around the X points $(\pi,0)$ and $(0,\pi)$.  The excellent agreement suggests that the \emph{Ansatz} in Eq. (\ref{eq:singlehole}) indeed accurately captures the quasiparticle degrees of freedom in the single-hole states. Consistent with our results, ARPES measurements on parent compounds of cuprate superconductors \cite{Ronning1998,Hu_2018} show that the lowest-energy states reside at the $\mathbf{k}_0$ momenta. However, the dispersion around these points is more isotropic in experiments, which can be attributed to a negative next-nearest-neighbor hopping $t'$ in real materials \cite{PhysRevB.51.8676,TakamiTohyama_2000}.

\subsection{Nontrivial loop current pattern}

The emergent nontrivial quantum number $L_z = \pm 1$ of the single-hole ground state implies the presence of non-zero currents. (Note that a unique ground state with angular momentum $L_z = 0$ or $L_z = 2$ exhibits no current by symmetry analysis.) Numerically the spin and charge currents in the single-hole ground state have been revealed by ED on a $4\times 4$ square lattice \cite{PhysRevB.98.165102}, which is further confirmed by DMRG at both $6\times 6$ and $8\times 8$ systems. Here the definitions of the hole and spin currents can be deduced \cite{PhysRevB.99.205128} from the continuity equations as follows:
\begin{eqnarray}
J_{i j}^h&=&i t \sum_\sigma\left(c_{i \sigma}^{\dagger} c_{j \sigma}-c_{j \sigma}^{\dagger} c_{i \sigma}\right),\label{eq:J_h}\\
J_{i j}^s&=&i\frac{J}{2}\left(S_i^{+} S_j^{-}-S_i^{-} S_j^{+}\right) ,\label{eq:J_s}\\
J_{i j}^b&=&-i \frac{t}{2} \sum_\sigma \sigma\left(c_{i \sigma}^{\dagger} c_{j \sigma}-c_{j \sigma}^{\dagger} c_{i \sigma}\right),\label{eq:J_b}
\end{eqnarray}
where $J^h$ denotes the hole current, $J^s$ represents the neutral spin current associated with the superexchange term $H_J$, and $J^b$ is the backflow spin current generated by the motion of the hole in $H_t$.

Figure~\ref{fig:currenttotal} displays the patterns of the hole current $J^h$ and total spin current $J^s_{\mathrm{tot}} = J^s + J^b$ on an $8\times 8$ lattice, calculated by both DMRG and VMC methods. Similarly, the results for $6\times 6$ system are presented in Appendix~\ref{app:secA}. One can see that the variational ground-state Eq.~(\ref{eq:singlehole}) can reproduce the detailed loop current patterns fairly consistent with the DMRG results. Here the hole currents form closed loops with a minimal size of $2 \times 2$ on the lattice, with the neighboring loops separated by distance $2a_0$ along the $\hat{x}$ and $\hat{y}$ directions. For the $8\times 8$ system, the strength of the hole current reaches its maximum within the central $4a_0 \times 4a_0$ area, where the hole density concentrates. For the $6\times 6$ system, a $2 \times 2$ loop current with a maximal strength is located on the central plaquette (see Fig.~\ref{fig:currentlx6}). The sensitivity of loop currents to the boundary conditions indicates a single hole can no longer be regarded as a point quasiparticle as will be further explored later, which is a new factor that complicates conventional DMRG studies of 2D doped Mott insulators as a small finite size is usually employed along the $\hat{y}$ direction in a ladder-like geometry in a DMRG simulation.

The presence of the loop current pattern exhibits remarkable robustness. For $t/J =3$, our VMC calculation confirms its existence for lattice sizes up to at least $14\times 14$.
Although the two-fold degeneracy of the single-hole ground state only persists over a finite range of $t/J$ \cite{PhysRevB.98.165102}, it remains robust even as the AFM long-range order is artificially suppressed with replacing $|\phi_0\rangle$ by a short-range dimer state in Eq.~(\ref{eq:singlehole}),  still yielding the $2 \times 2$ loop currents within a specific $t/J$ range. Moreover, the previous DMRG studies also confirm its survival in 3- and 5-hole doped systems ($6\times 6$ lattice) \cite{PhysRevB.98.165102}, suggesting its stability in finite-doping regimes with an odd number of holes across varying spin correlation lengths.

\subsection{Emergent magnetic moment}

\begin{figure}[h]
\includegraphics[width=0.50\textwidth]{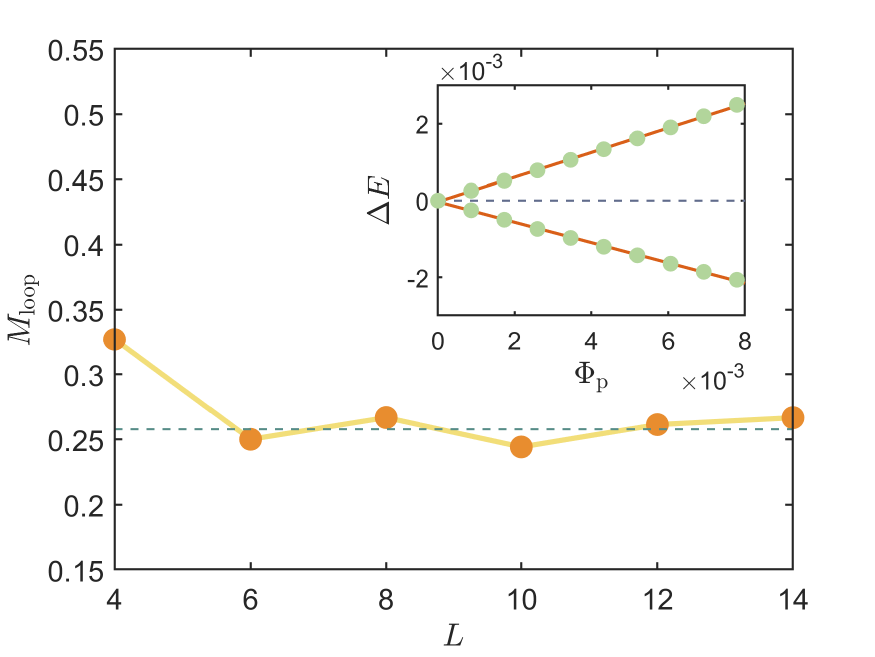}% Here is how to import EPS art
\caption{\label{fig:moment_scaling} Magnetic moment $M_{\mathrm{loop}}$ of the loop current as a function of system size $L$. Inset: the energy changes, $\Delta E = E(\Phi_{\mathrm{p}}) - E(\Phi_{\mathrm{p}} =0)$, induced by the polarization of $M_{\mathrm{loop}}$ via magnetic flux $\Phi_{\mathrm{p}}$ on a $12\times 12$ system. }
\end{figure}

The $2 \times 2$ loop current pattern in the single-hole ground state will give rise to an emergent magnetic moment pointing perpendicular to the 2D plane. An external magnetic field lifts the ground-state degeneracy, selecting the state whose magnetic moment $\vec{M}_{\mathrm{loop}}$ aligns with the field $\vec{B}$ with an energy shift $\Delta E = -\vec{M}_{\mathrm{loop}} \cdot \vec{B}$. As shown in the inset of Fig.~\ref{fig:moment_scaling}, we introduce a uniform magnetic flux via Peierls substitution $-t c_{i\sigma}^{\dagger} c_{j\sigma} \rightarrow -t e^{i A^e_{ij}} c_{i\sigma}^{\dagger} c_{j\sigma}$ in the hopping term Eq.~(\ref{eq:Ht}), and compute the energy splitting of the two lowest states with opposite magnetizations as a function of flux per plaquette $\Phi_{\mathrm{p}}$. The slope of the $\Delta E$ – $\Phi_{\mathrm{p}}$ curve yields the magnetization ${M}_{\mathrm{loop}}$ contributed by the loop current. Figure~\ref{fig:moment_scaling} shows that $M_{\mathrm{loop}}$ quickly converges to a finite value of $0.22$ for system sizes larger than the intrinsic loop current pattern of $4\times 4$. Inserting the unit of $M_{\mathrm{loop}}$, $Jea_0^2/\hbar$, and using realistic cuprate parameters — the lattice constant $a_0 = 4~\text{\AA}$ and superexchange coupling constant $J = 0.1~\mathrm{eV}$ — we estimate $M_{\mathrm{loop}} \approx 0.1~\mu_B$, which is detectable experimentally \cite{PhysRevLett.96.197001,PhysRevLett.89.247003}.

\begin{figure*}[t]
\includegraphics[width=0.65\textwidth]{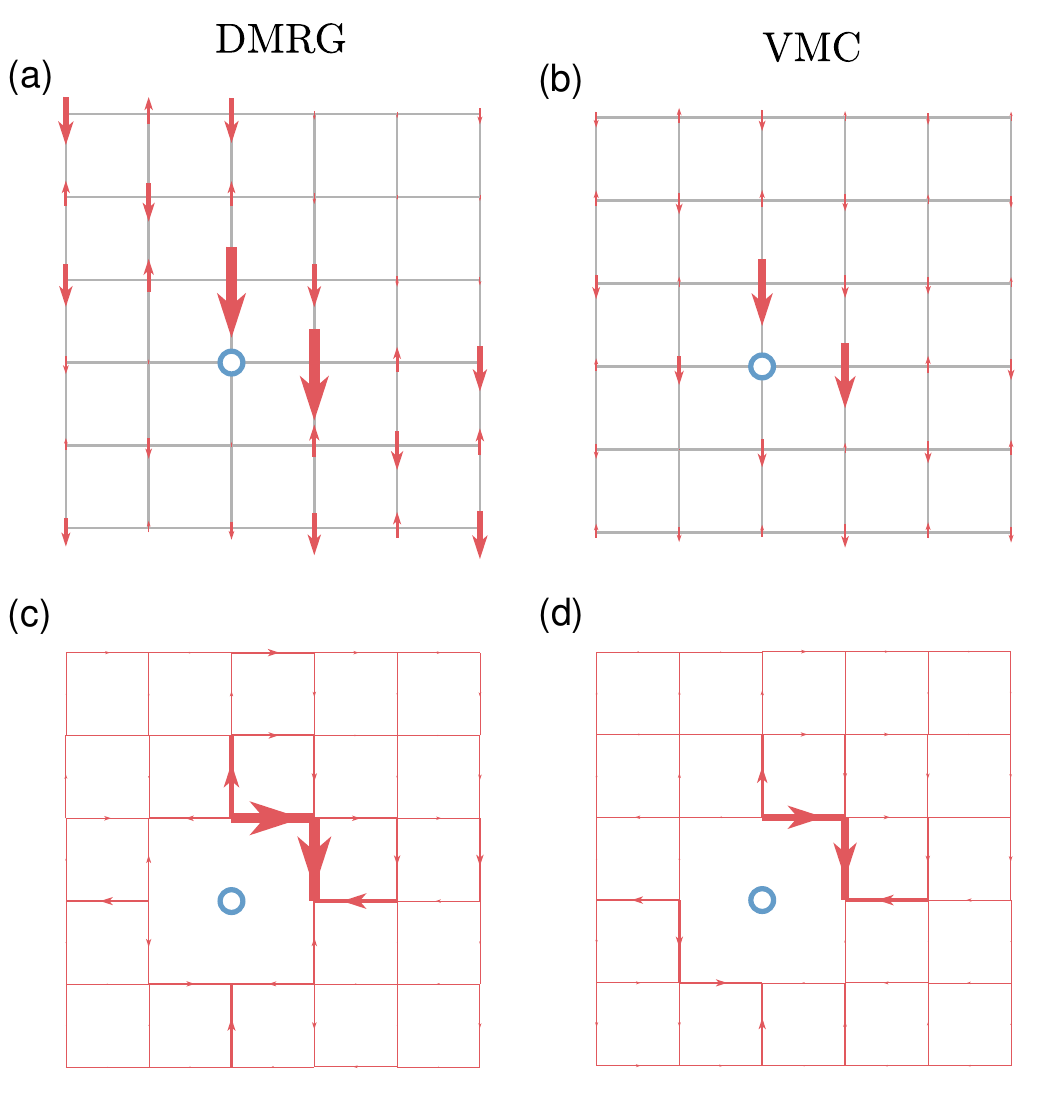}% Here is how to import EPS art
\caption{\label{fig:nhSizPJs} 
(a, b) Hole–spin correlator $\langle n^h_{i} S^z_{j}\rangle$ and (c, d) neutral spin current pattern 
$\langle n^h_{i} J^s_{jk}\rangle$ for the degenerate ground state with quantum numbers $L_z = 1$, $S^z = -\frac{1}{2}$. The projected position of the hole $i$ is marked by a blue circle, and the results obtained with DMRG (left column) and VMC (right column) are comparatively shown. In (a, b), the direction and length of each arrow represent the sign and magnitude of the correlator. In (c, d), arrow thickness indicates the relative magnitude of the current. Panels (a, b) share a common scale, and panels (c, d) share a separate common scale. Note that the hole position is slightly off the center of the $6\times 6$ lattice, such that the $S^z = -\frac{1}{2}$ and spin current surrounding the hole look asymmetric. } 
\end{figure*}

\subsection{The hole state as a charge-spin composite}

Finally, let us examine the origin of spin loop currents in Figs. \ref{fig:currenttotal}(c) and (d). The DMRG and VMC calculations of the hole-spin correlator $\langle n^h_{i} S^z_{j}\rangle$ both reveal an enhanced density of $\downarrow$ spins around the hole projected at a site shown in Figs.~\ref{fig:nhSizPJs} (a) and (b) for an $S^z=-\frac{1}{2}$ ground state [with $\sigma=\uparrow$ in Eq.~(\ref{eq:singlehole})]. The local magnetization $\langle S^z_j\rangle$ around the hole creates a slightly more ferromagnetic spin environment to facilitate the hole hopping, which favors loose binding of the spin-1/2 partner around the hole despite the long-range spin-spin correlation in the background. Furthermore, the accumulated $\downarrow$ spins (total $S^z=-\frac{1}{2}$) are actually circulating around the hole as illustrated by Figs. \ref{fig:nhSizPJs}(c) and (d), indicating that once the charge and spin are partially separated, they immediately produce mutual currents around each other to form a hole composite. This process is strongly generated by the hopping term, as schematically illustrated in Fig.~\ref{fig:string}(b), which will be further elaborated in the following section.

\section{Single-hole wave function: A quantum ``cat state''}\label{sec:3}

In the last section, we have shown that a single hole injected into a Heisenberg AFM spin background can be well described by the variational \emph{Ansatz} Eq.~(\ref{eq:singlehole}), which predicts a dual behavior in excellent agreement with the large-scale numerical simulations: On one hand, the single hole exhibits a quasiparticle-like dispersion extending over the whole Brillouin zone; on the other hand, it shows minimal loop current patterns at the plaquette scale, which seems incompatible with a point-like Bloch quasiparticle description. As the unified wave function, Eq.~(\ref{eq:singlehole}) will be carefully analyzed in the following to reveal the novel nature of strong correlation that leads to such a dual picture.    

\subsection{Resonating dual components of the single-hole state}

The loop current pattern in Fig.~\ref{fig:currenttotal} implies the presence of an incoherent component in  the single-hole ground state. This observation motivates decomposing the wave function Eq.~(\ref{eq:singlehole}) in the following form:
\begin{equation}
    |\Psi_{\mathrm{G}}\rangle_{1\mathrm{h}} = |\Psi_{\text{qp}}\rangle_{1\mathrm{h}} + |\Psi_{\text{inc}}\rangle_{1\mathrm{h}},
    \label{eq:1h_component}
\end{equation}
where $|\Psi_{\text{qp}}\rangle_{1\mathrm{h}}$ takes the bare hole form $\sum_i\varphi(i)c_{i\sigma} |\phi_0\rangle$ and its norm is associated with the quasiparticle spectral weight in Eq.~(\ref{eq:specweight_def}) via $_{1\mathrm{h}}\langle \Psi_{\text{qp}}|\Psi_{\text{qp}}\rangle_{1\mathrm{h}} \equiv 2 \sum_{\mathbf{k}} Z_{\mathbf{k}}$, by noting that the incoherent component $|\Psi_{\text{inc}}\rangle_{1\mathrm{h}}$ is orthogonal to the quasiparticle component: $_{1\mathrm{h}}\langle \Psi_{\text{qp}} | \Psi_{\text{inc}} \rangle_{1\mathrm{h}} = 0$ and $\varphi(i) = 2\langle \phi_0 | c_{i\sigma}^{\dagger} |\Psi_{\mathrm{G}}\rangle_{1\mathrm{h}}$ with the factor $2$ coming from the electrons being half-filling in $|\phi_0\rangle$. Physically, the $|\Psi_{\text{qp}}\rangle_{1\mathrm{h}}$ and $|\Psi_{\text{inc}}\rangle_{1\mathrm{h}}$ components represent the average part and fluctuation parts of the phase-shift operator $e^{-im\left(\hat{\Omega}_{i}-\hat{\Omega}_{v}\right)}$ in single-hole wave function Eq.~(\ref{eq:singlehole}), as demonstrated in Appendix \ref{app:fluctuation}. 

\begin{table}[h]
\caption{\label{tab:table1}
Energy contributions of the quasiparticle and incoherent components for the single‑hole ground state on an $8\times8$ lattice, calculated using VMC. All values are given in units of $J$. The hopping and superexchange terms are denoted by $\hat{H}_{t}$ and $\hat{H}_J$, respectively, as defined in Eqs.~(\ref{eq:Ht}) and (\ref{eq:HJ}); $\hat{H}_{t\text{-}J}$ is the full Hamiltonian. 
$E_0$ refers to the superexchange energy of the half‑filled (undoped) system.
}
\begin{ruledtabular}
\centering
\begin{tabular}{cccc}
\multicolumn{1}{c}{$\hat{O}=$} &
\multicolumn{1}{c}{$\hat{H}_{t\text{-}J}-E_0$} &
\multicolumn{1}{c}{$\hat{H}_t$} &
\multicolumn{1}{c}{$\hat{H}_J-E_0$} \\   
\colrule
$_{1\mathrm{h}}\langle \Psi_{\text{qp}}|\hat{O}|\Psi_{\text{qp}}\rangle_{1\mathrm{h}}$ & $0.915$ & $-0.038$ & $0.953$ \\ 
$ _{1\mathrm{h}}\langle \Psi_{\text{inc}}|\hat{O}|\Psi_{\text{inc}}\rangle_{1\mathrm{h}}$ & $0.476$ & $-1.470$ & $1.946$\\ 
$ _{1\mathrm{h}}\langle \Psi_{\text{qp}}|\hat{O}|\Psi_{\text{inc}}\rangle_{1\mathrm{h}} +h.c.$ & $-5.243$ & $-5.152$ &  $-0.092$ \\ 
$_{1\mathrm{h}}\langle \Psi_{\mathrm {G}}|\hat{O}|\Psi_{\mathrm{G}}\rangle_{1\mathrm{h}}$ & $-3.853$ & $-6.659$  & $2.807$ \\ 
\end{tabular}
\end{ruledtabular}
\end{table}

The energy contributions from $|\Psi_{\text{qp}}\rangle_{1\mathrm{h}}$ and $|\Psi_{\text{inc}}\rangle_{1\mathrm{h}}$, along with their cross terms, are summarized in Table~\ref{tab:table1} for the $8 \times 8$ lattice size. It reveals that the kinetic energy contribution from the quasiparticle component is actually very small ($\sim -0.038$), which stems from the fact that the dominant Bloch-wave component at $\mathbf{k}_0= (\pm\frac{\pi}{2},\pm\frac{\pi}{2})$ vanishes in a tight-binding model: $-2t_{\text{eff}}(\cos(k_{0x})+\cos(k_{0y})) = 0$. In contrast, the kinetic energy contribution from $|\Psi_{\text{inc}}\rangle_{1\mathrm{h}}$ is much larger ($\sim -1.470$), although it is 
offset by a substantial superexchange energy cost ($\sim 1.946$).
Totally, the dominant kinetic energy gain ($\sim -5.152$) comes from the off-diagonal transition between $|\Psi_{\text{qp}}\rangle_{1\mathrm{h}}$ and $|\Psi_{\text{inc}}\rangle_{1\mathrm{h}}$. In other words, the off-diagonal contribution to the single-hole ground-state energy is much larger than the difference between the two diagonal contributions. Correspondingly, the calculated $\langle \Psi_{\text{qp}}|\Psi_{\text{qp}}\rangle_{1\mathrm{h}}= 0.447$ and $\langle \Psi_{\text{inc}}|\Psi_{\text{inc}}\rangle_{1\mathrm{h}}= 0.553$ are quite close. It means the single-hole ground state in Eq.~(\ref{eq:1h_component}) forms a ``cat state'' with the two components strongly resonating, which cannot be approached perturbatively from the quasiparticle side. 

Indeed, if one starts from the bare hole state $|\Psi_{\text{qp}}\rangle_{1\mathrm{h}}$ alone,
even with incorporating the ``longitudinal spin polaron effects" via, e.g., a first-order Lanczos step, the VMC calculation fails to yield the nontrivial quantum number $L_z = \pm 1$ or the spin/hole current patterns observed in DMRG calculations (See Appendix \ref{app:longitudinal} for details).  
The ``hidden" incoherent part $|\Psi_{1\mathrm{h}}\rangle_{\text{inc}}$ plays a pivotal role in determining the physical properties of the single-hole ground state. Here, the quasiparticle is emergent as a component of Eq.~(\ref{eq:singlehole}) rather than the fundamental building block, implying a critical departure from the conventional Landau Fermi liquid paradigm.

Physically, the creator of the bare hole $c_{i\sigma}=\tilde{c}_{i\sigma}e^{\pm i\hat{\Omega}_{i}}$ may be viewed as originating from the tight bonding of the twisted hole $\tilde{c}_{i\sigma}$ with the excitation of antimeron $e^{-im\hat{\Omega}_{v}}$ at $v\rightarrow i$. Thus, the quasiparticle component $|\Psi_{\text{qp}}\rangle_{1\mathrm{h}}$ is primarily contributed in Eq.~(\ref{eq:singlehole}) by the hole-antimeron pair with small $|i-v|$. Correspondingly, the average part of the phase-twist operator $\langle e^{\mp i(\hat{\Omega}_i-\hat{\Omega}_v)}\rangle$, which governs the quasiparticle component discussed above, decays exponentially with distance $|i-v|$ \cite{PhysRevB.111.104502}. In the ground state, $|\Psi_{\text{qp}}\rangle_{1\mathrm{h}}$ is basically a Bloch-wave at momenta $\mathbf{k}_0= (\pm\frac{\pi}{2},\pm\frac{\pi}{2})$ [cf. Fig.~\ref{fig:Akwneg}], four of which are combined into a state with angular momentum $L_z = \pm 1$ (cf. Eq.~(\ref{eq:Lz1quasi}) in Appendix \ref{app:phase} for details). Moreover, since $|\phi_0\rangle$ as a spin singlet ground state does not break the translational symmetry even in the large-sample limit, the four peak momenta $\mathbf{k}_0= (\pm\frac{\pi}{2},\pm\frac{\pi}{2})$ are unrelated to the usual AFM folding effect. As a matter of fact, variationally $|\phi_0\rangle$ may be tuned into a short-range AFM state, where the equal energy contours around four $\mathbf{k}_0$ are still maintained. 

On the other hand, the incoherent component $|\Psi_{\text{inc}}\rangle_{1\mathrm{h}}$ emerges in Eq.~(\ref{eq:1h_component}) once the twisted hole and the antimeron are spatially separated such that the hole state in Eq.~(\ref{eq:singlehole}) becomes a composite object. Here, the chiral spin current vortex surrounding the twisted hole can no longer be effectively compensated by a tight-binding antimeron, and a transverse distortion induced by the dipole-like phase-shift operator $e^{\mp i(\hat{\Omega}_{i}-\hat{\Omega}_{v})}$ intensifies. This distortion corresponds to the ``transverse spin-polaron effect'', which is manifested in the local spin-current pattern around the hole shown in Figs. \ref{fig:nhSizPJs} (c)(d). The consistency between VMC and DMRG implies the \emph{Ansatz} accurately captures this effect.

\subsection{The antimeron}

\begin{figure}[h]
\includegraphics[width=0.45\textwidth]{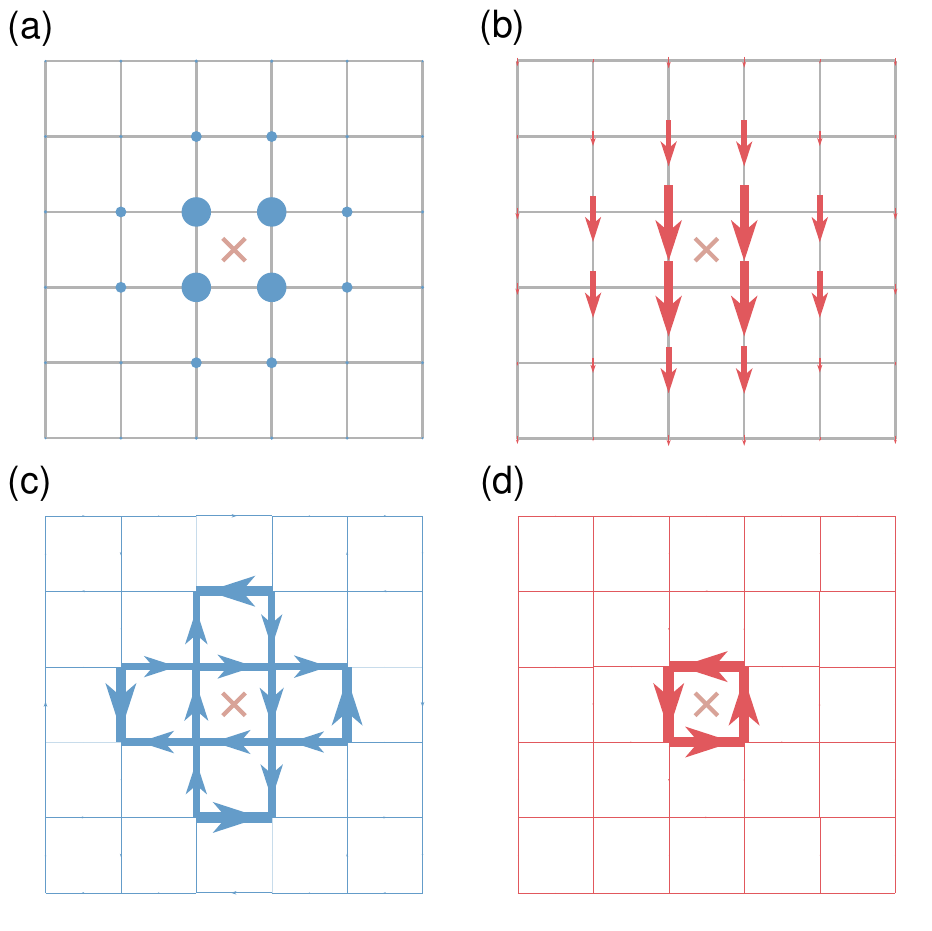}% Here is how to import EPS art
\caption{\label{fig:antimeronproj} 
Real-space profiles for the single-hole ground state with an antimeron pinned at the center of a plaquette $v_0$ (marked by the pink cross) in a $6\times 6$ lattice. Shown are (a) the hole density 
$n_i^h$, (b) the spin density $S^z_i$, (c) the charge current pattern, and (d) the total spin current distribution. These quantities illustrate the composite structure of the hole bound to a topological defect in the spin background.} 
\end{figure}

Let us further examine the incoherent component in the single-hole ground state.  In contrast to a point-like Bloch-wave quasiparticle usually defined on lattice sites, we may fix the antimeron at position $v$ and analyze the corresponding wave function in Eq.~(\ref{eq:singlehole}):
\begin{equation}
    |\Psi_{\mathrm{G}}(v)\rangle_{1\mathrm{h}} \equiv \sum_{i,m=\pm 1}\varphi_{m}(i,v)c_{i\sigma}e^{-im\left(\hat{\Omega}_{i}-\hat{\Omega}_{v}\right)}|\phi_{0}\rangle. \label{eq:projantimeron}
\end{equation}
Fig.~\ref{fig:antimeronproj}(a) displays the hole density distribution with the antimeron projected at the center of a plaquette labeled by the cross, i.e., $_{1\mathrm{h}}\langle \Psi_{\mathrm{G}}(v_0)|n^{h}_i|\Psi_{\mathrm{G}}(v_0)\rangle_{1\mathrm{h}}$. The hole density is predominantly localized on the four nearest-neighbor sites of the projected antimeron, forming the $2 \times 2$ structure that confirms a strong attractive interaction between the hole and the antimeron. A spin-1/2 (here $\sigma=\uparrow$ is considered such that $S^z=-1/2$) is also distributed slightly loosely around Fig.~\ref{fig:antimeronproj}(b). Furthermore, Figs.~\ref{fig:antimeronproj}(c) and \ref{fig:antimeronproj}(d) show the corresponding charge and spin currents surrounding $v_0$ for $L_z=+1$ ground state. Here, the local non-quasiparticle patterns, where the doped hole and its spin are closely distributed around $v_0$ and form the charge and spin loop currents, are clearly specified by the antimeron coordinate $v_0$, which is fixed at the center of the sample $6 \times 6$. 

\begin{figure}[h]
\includegraphics[width=0.35\textwidth]{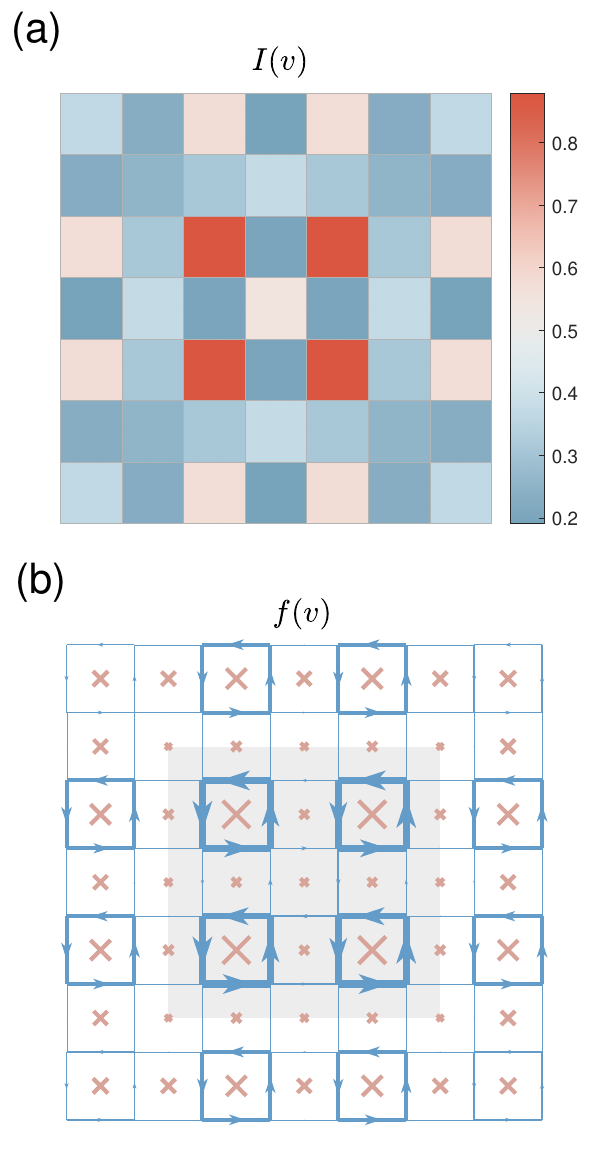}
\caption{\label{fig:antimeron} (a) Spatial distribution of the ratio $I(v)$ [Eq.~(\ref{eq:Iv})] on each plaquette $v$ in the $8\times 8$ lattice. (b) Antimeron strength $f(v)$ [Eq.~(\ref{eq:fv})] and VMC-calculated hole current $J^h$ in the $8\times 8$ lattice. The size of each cross is proportional to $f(v)$. The gray-colored square in the central region marks an emerging basic $4a_0 \times 4a_0$ structural unit in a large sample size.}  
\end{figure}

Note that the detailed spin-charge patterns can vary with the antimeron locations $v$. The quasiparticle component of $|\Psi_{\mathrm{G}}(v_0)\rangle_{1\mathrm{h}}$ is also mixed here but its weight remains very weak when $v_0$ coincides with the loop-current center of the $6\times6$ sample [cf. Fig.~\ref{fig:currentlx6}]. To quantify this behavior, we decompose $|\Psi_{\mathrm{G}}(v)\rangle_{1\mathrm{h}}$ in Eq.~(\ref{eq:projantimeron}) into quasiparticle component $|\Psi_{\mathrm{qp}}(v)\rangle_{1\mathrm{h}}$ and incoherent component $|\Psi_{\mathrm{inc}}(v)\rangle_{1\mathrm{h}}$. We define the ratio:
\begin{equation}I(v)  = \frac{ _{1\mathrm{h}}\langle \Psi_{\text{inc}}(v)|\Psi_{\text{inc}}(v)\rangle_{1\mathrm{h}}} { _{1\mathrm{h}}\langle \Psi_G(v)|\Psi_G(v)\rangle_{1\mathrm{h}} }, \label{eq:Iv}
\end{equation}
which measures the relative weight of the incoherent component on each plaquette $v$. The calculated spatial distribution of $I(v)$ on an $8\times8$ lattice is shown in Fig.~\ref{fig:antimeron}(a).

A pronounced spatial heterogeneity is clearly revealed:  single-hole ground state is not a uniform mix of quasiparticle and incoherent component. Instead, certain plaquettes are dominated by the incoherent component (large $I(v)$), while others are dominated by the quasiparticle component (small $I(v)$). Most importantly, by comparing this distribution with the hole current pattern obtained from VMC [Fig.~\ref{fig:antimeron}(b)] (or DMRG results in Fig.~\ref{fig:currenttotal}(a)), we find that the centers of the macroscopic $2\times 2$ loop currents coincide exactly with the plaquettes $\{v_j\}$ where $|\Psi_{\mathrm{G}}(v_j)\rangle_{1\mathrm{h}}$ is dominated by the incoherent component. This correlation is independent of system size, reinforcing the physical picture that the loop currents are direct manifestations of the incoherent hole-antimeron bound state, rather than a byproduct of the Bloch quasiparticle.

The staggered spatial pattern of quasiparticle and incoherent components is further correlated with the antimeron strength
\begin{equation}\label{eq:fv}
f(v)=\sum_{m=\pm 1,i}|\varphi_m(i,v)|^2,
\end{equation}
which reflects the background spin texture induced by the ``feedback effect'' of the doped hole. The distribution of $f(v)$ for the $8\times8$ system is shown in Fig.~\ref{fig:antimeron}(b). Here, $f(v)$ is maximized at centers of the hole current loops, with a $4a_0 \times 4a_0$ checkerboard pattern, as indicated by the dashed square,  emerging in the $8\times 8$ and larger systems, whereas $f(v)$ is maximized at the single $2 \times 2$ charge loop current appears at the center of the sample $6 \times 6$ [cf. Fig.~\ref{fig:currentlx6}]. The minimal periodicity is $4a_0 \times 4a_0$ rather than $2a_0 \times 2a_0$ because there is a $\pi$ phase shift of the wave function upon $2a_0$ translation along the $\hat{x}$ and $\hat{y}$ directions (See Appendix ~\ref{app:phase} for details).

Before we further discuss this spatial-alternating property of the two components on the lattice in the following, let us reduce the system to a minimal $2\times 2$ lattice to further examine the origin of the loop currents.

\subsection{The doped hole as a ``closed string''}
\begin{figure}[h]
\includegraphics[width=0.48\textwidth]{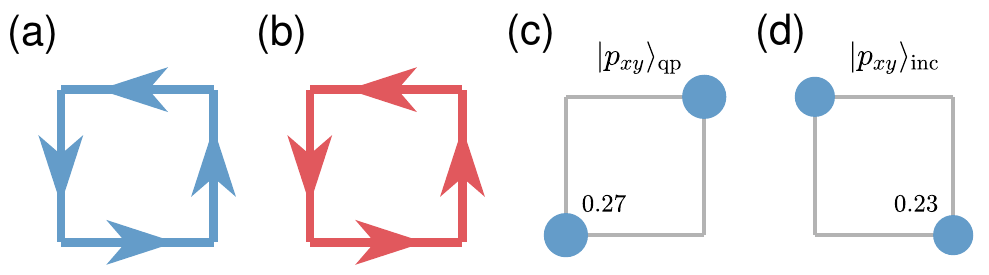}% Here is how to import EPS art
\caption{\label{fig:currentLz1ED} (a) Hole current $J^h$ and (b) total spin current $J^s_{\mathrm{tot}}$ for the ground state with $S^z = -\frac{1}{2}$ and $L_z = 1$ on a $2\times 2 $ lattice. The corresponding hole density distributions of the quasiparticle component $|p_{xy}\rangle_{\text{qp}}$ and incoherent component $|p_{xy}\rangle_{\text{inc}}$ of the real wave function [defined in Eq.~(\ref{eq:realpwave})] are displayed in panels (c) and (d), respectively.}
\end{figure}

Previously, we have seen that the doped hole is tightly bound to an antimeron at the four sites of a plaquette, in which the antimeron is centered. It is suggested that the minimal size for the incoherent component of the doped hole to appear may be a four-site chain with periodic boundary condition (PBC). In the following, we compare the VMC result with ED for such a 4-site system to elucidate the underlying local physics.    

Here, the ground state degeneracy depends critically on the ratio $t/J$, reflecting the competition between the kinetic term and superexchange term (detailed analysis is provided in Appendix \ref{app:ED}). For $0.5<t/J<3.8$, the double degeneracy with $L_z=\pm 1$ for the $2\times 2$ plaquette indeed emerges (at a given $S^z=\pm 1/2$), which corresponds to momenta $k=\pm\frac{\pi}{2}$ in the 1D picture. The loop current patterns of the charge and spin at $t/J=3$ are shown in Figs.~\ref{fig:currentLz1ED}(a) and \ref{fig:currentLz1ED}(b), respectively, with a good comparison with the VMC result (not shown). Furthermore, one may also construct two degenerate real wave functions, i.e., the $p$-wave $|p_{xy}\rangle$ and $|p_{yx}\rangle$ based on $L_z=\pm 1$ states (cf. Eq.(\ref{eq:realpwave}) below). In Figs.~\ref{fig:currentLz1ED}(c) and \ref{fig:currentLz1ED}(d), the hole densities of the state $|p_{xy}\rangle$ for the quasiparticle and incoherent components are illustrated, respectively. Later in the next subsection, we will further discuss the one-hole ``cat state'' based on the real wave function at larger 2D systems.   

Therefore, Fig.~\ref{fig:currentLz1ED} for the $2\times 2$ system reproduces the $2 \times 2$ loop current of the hole in a larger system (cf. Fig.~\ref{fig:antimeronproj} for the $8\times 8$ system), which is approximately a minimal object of the hole motion along a 4-site ring at $k=\pm\frac{\pi}{2}$ as a ``closed string". This emergent string-like excitation — carrying the internal angular momentum $L_z$ — deviates fundamentally from a point-like quasiparticle, challenging the single-particle paradigm of Landau Fermi liquid theory. 

Finally, in Appendix \ref{app:phase}, it is shown that an approximate Bloch-wave-like state at $\mathbf{k}_0= (\pm\frac{\pi}{2},\pm\frac{\pi}{2})$ may be constructed based on the $2 \times 2$ ``closed string'' structures with $(\frac{\pi}{2},\frac{\pi}{2})$ momentum in a reduced Brillouin zone.
While the plaquette-based effective theories \cite{PhysRevB.76.161104,PhysRevB.65.104508,PhysRevB.66.180503} treat the plaquette as a fundamental unit, we point out that the doped hole as a $2\times 2$ closed string may not be a rigid building block in constructing the 2D eigenstates, but rather a dynamically emergent structure arising from the local hole motion.

\begin{figure}[t]
\includegraphics[width=0.45\textwidth]{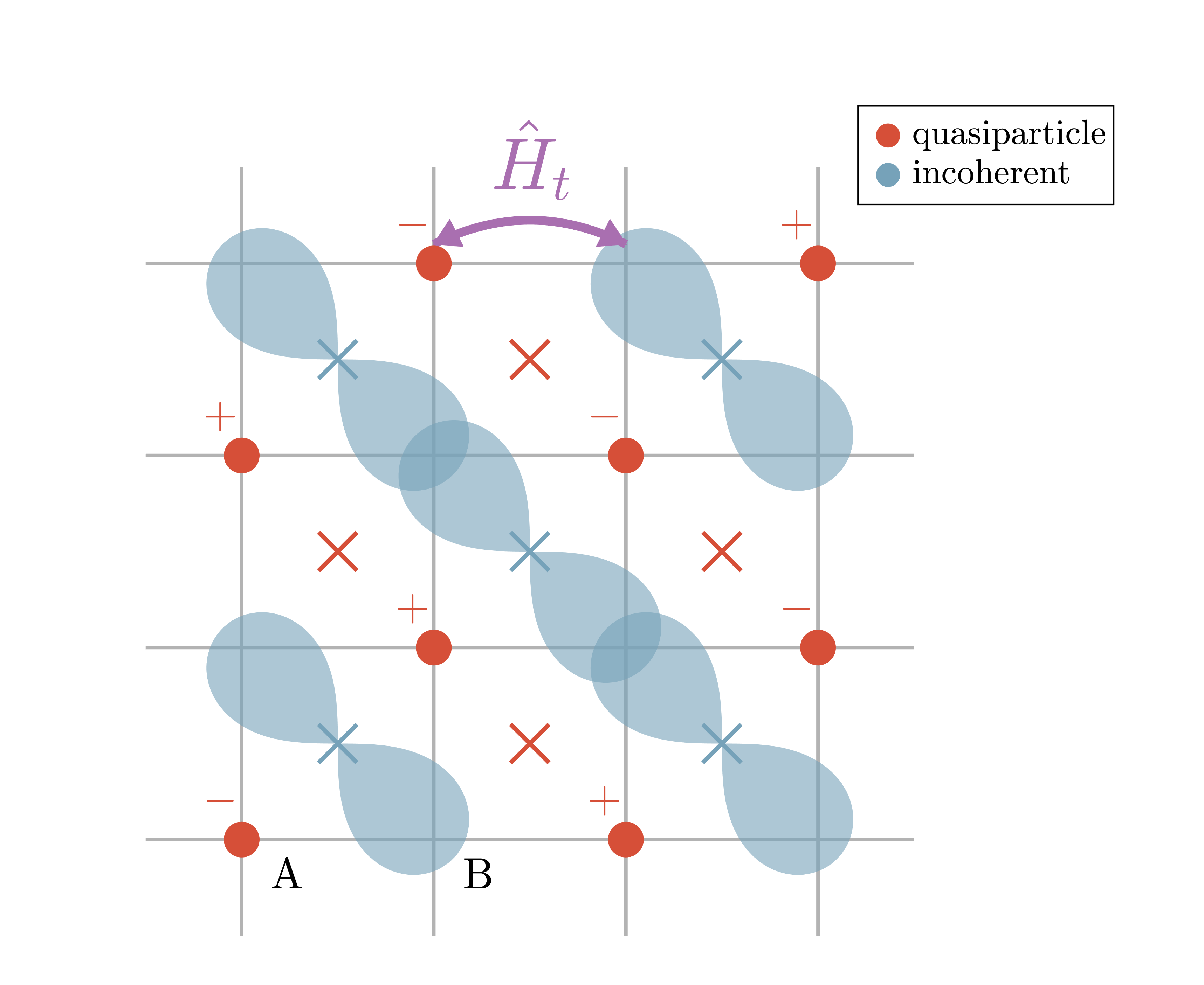}
\caption{\label{fig:pwave} Red and blue crosses mark plaquette positions $v$ where $|p_{xy}(v)\rangle$ is dominated by the quasiparticle or incoherent component, respectively; the $\pm$ labels indicate the phase of $\varphi(i)$ in the quasiparticle expansion 
 $\sum_i \varphi(i) c_{i\uparrow} |\phi_0\rangle$. The spindle-shaped blue orbitals represent the emergent $p$-wave character of the incoherent component when an antimeron is projected onto the sites marked by blue crosses. Notably, the quasiparticle component $|p_{xy}\rangle_{\text{qp}}$
has nearly zero weight on the $B$ sublattice. For clarity, the small weights of $|p_{xy}\rangle_{\text{inc}}$ on the $A$ sublattice are omitted from the plot.}
\end{figure}

\subsection{$p$-wave orbital representation}

The single-hole ground state [Eq. (\ref{eq:singlehole})] with $L_z =\pm 1$ may be reexpressed as real wave functions as follows:
\begin{equation}
    \begin{aligned}
        |p_{xy}\rangle &= \frac{1}{\sqrt{2}}(|L_z = -1\rangle + |L_z = +1\rangle), \\
        |p_{yx}\rangle &= \frac{-i}{\sqrt{2}}(|L_z = -1\rangle - |L_z = +1\rangle).
    \end{aligned} \label{eq:realpwave}
\end{equation}
The states $|p_{xy}\rangle$ and $|p_{yx}\rangle$ exhibit no current and are of $p$-wave symmetry. Here a specific phase convention for $L_z=\pm1$ states is chosen, under which the Bloch-wave component of $|p_{xy}\rangle$ appears only at momenta $(\pi/2,\pi/2)$ and $(-\pi/2,-\pi/2)$, and that of $|p_{yx}\rangle$ at $(\pi/2,-\pi/2)$ and $(-\pi/2,\pi/2)$. In the following, we focus solely on the $|p_{xy}\rangle$ state, as $|p_{yx}\rangle$ state can be obtained by rotating $|p_{xy}\rangle$ by $\pi/2$. Following the same procedure as in Eq.~(\ref{eq:1h_component}), we decompose $|p_{xy}\rangle$ into the quasiparticle and incoherent components, $|p_{xy}\rangle_{\text{qp}}$ and $|p_{xy}\rangle_{\text{inc}}$, respectively, and measure their respective hole density distributions:  $_{\text{qp}}\langle p_{xy}|n^h_i|p_{xy}\rangle_{\text{qp}}$ and $_{\text{inc}}\langle p_{xy}|n^h_i|p_{xy}\rangle_{\text{inc}}$, as previously shown in Figs.~\ref{fig:currentLz1ED}(c) and \ref{fig:currentLz1ED}(d) for the 4-site ring case. 

As shown in Fig.~\ref{fig:pwave}, for the $8\times 8$ system in the central region, one finds that the hole densities corresponding to the two components are spatially separated: one is predominantly distributed on the $A$ sublattice, while the other is mostly on the $B$ sublattice (cf. Fig.~\ref{fig:qp_incoh_lx8} in Appendix~\ref{app:pwave}). In this staggered pattern, the spatial separation of the two components at nearest neighboring sublattices is connected by the hopping term $\hat{H}_t$ as illustrated in Fig.~\ref{fig:pwave}. The kinetic energy of the hole predominantly originates from the resonance between the quasiparticle and incoherent components similar to the previous complex wave functions. 

The real-space pattern in Fig.~\ref{fig:pwave} depends critically on the antimeron position $v$  marked by crosses at the center of each plaquette of the square lattice, confirming $v$ as an emergent dynamic coordinate. Projecting the antimeron at a specific site $v$ as in Eq.~(\ref{eq:projantimeron}) will yield $|p_{xy}(v)\rangle$ in the present case. Alternating at the dual lattice, $v$'s are marked by red and blue in Fig.~\ref{fig:pwave}, which indicates $|p_{xy}(v)\rangle$ is of either predominately the quasiparticle component (cf. Fig.~\ref{fig:qp_dominate_lx8} in Appendix~\ref{app:pwave}) or the incoherent $p$-wave-like orbital (cf. Fig.~\ref{fig:incoh_dominate_lx8} in Appendix~\ref{app:pwave}). 
Note that here additional hole density with minor weight outside the orbital is omitted in Fig.~\ref{fig:pwave} for clarity.  
The sign change characteristic of the $p$-wave orbital is confirmed by $\varphi_m(i_1, v) \approx -\varphi_m(i_2, v)$, where $i_1$ and $i_2$ denote two lobes of the orbital. The blue $v$ positions, at which $|p_{xy}(v)\rangle$ is dominated by the incoherent component, coincide with the centers of the $2\times 2$ loop currents for the complex wave function in Fig.~\ref{fig:antimeron}.

\section{Pairing of two holes in a ``cat state''}\label{sec:4}

In the previous section, it has been shown that a single hole moving in an AFM background forms a quantum ``cat state''—a superposition of a bare hole component and an incoherent component. The essential non-perturbative strong correlation arises because the hole’s kinetic energy is gained mainly through resonance between these two distinct components: a quasiparticle part and a loop-current part, which constitute a non-Landau-type quasiparticle. When two holes are introduced, however, the structure of the state changes qualitatively. For the two-hole case, we demonstrate below that the ground state is also a quantum ``cat state'', now composed of a bare Cooper pair and an incoherent pair component with strong resonance between them. Here, the two holes are tightly bound at a scale much shorter than the long-range AFM correlation length, indicating that the pairing mechanism does not rely on long-wavelength spin-wave fluctuations. It is no longer a Cooper pair formed by the Landau quasiparticles. Instead, the key incoherent pair component emerges from the “fusion'' of the loop-current contributions from each individual hole, which constitutes the foundation of the proposed pairing mechanism.

\subsection{Two-hole ground-state wave function \emph{Ansatz}}

The variational wave function in Eq.~(\ref{eq:singlehole}) can be interpreted as a composite entity in which a doped hole is ``twisted'' and then is tightly bound to an antimeron texture in the AFM spin background. Introducing a second hole into the system naturally favors its binding with the latter antimeron texture, leading to a reduction in total energy. A paired state of two holes can thus be formulated as a natural generalization of the single-hole wave function in Eq.~(\ref{eq:singlehole}). A minimal \emph{Ansatz} for the two-hole ground state is given by:
\begin{equation}
|\Psi_{\mathrm{G}}\rangle_{2\mathrm{h}}=\sum_{i,j,m=\pm 1}g_{m}(i,j)c_{i\uparrow}c_{j\downarrow}e^{-im\left(\hat{\Omega}_{i}-\hat{\Omega}_{j}\right)}|\phi_{0}\rangle, \label{eq:twohole}
\end{equation} 
As mentioned above, this construction is analogous to adding an extra hole at the center of an antimeron in the single-hole wave function in Eq.~(\ref{eq:singlehole}), with the vortex center 
$v$ now located at the original lattice site $j$.

The ansatz in Eq. (\ref{eq:twohole}) has previously been studied by VMC in Ref. \cite{PhysRevX.12.011062}, where a dichotomy between a $d$-wave symmetry in the $c$-representation and a robust $s$-wave in terms of the twisted holes $\tilde{c}$ [i.e., the symmetry of $g_{m}(i,j)$] is demonstrated. In the following, we shall only focus on the $d$-wave symmetry in the $c$-representation, and show that it is a strong resonating ``cat state'' between a $d_{x^2-y^2}$-wave Cooper-pair channel and an incoherent $d_{xy}$ pairing channel. In other words, the two-hole pairing wave function indicates a new non-Cooper-pairing mechanism, which further suggests that a non-BCS-like SC state may emerge in the lightly-doped Mott insulator.

\subsection{Two distinct components of the two-hole wave function}

Analogous to the single-hole case, one may decompose the two-hole wave function into two components:
\begin{equation}
    |\Psi_{\mathrm{G}}\rangle_{2\mathrm{h}} = |\Psi_{\mathrm{Cooper}}\rangle_{2\mathrm{h}} + |\Psi_{\text{inc}}\rangle_{2\mathrm{h}}~,
    \label{eq:2h_component}
\end{equation}
where the quasiparticle/Cooper-pair component $|\Psi_{\mathrm{Cooper}}\rangle_{2\mathrm{h}}$ takes the conventional form $\sum_{i,j}\varphi(i,j)c_{i\uparrow}c_{j\downarrow} |\phi_0\rangle$. The incoherent component $|\Psi_{\text{inc}}\rangle_{2\mathrm{h}}$ is constructed to be orthogonal to $|\Psi_{\mathrm{Cooper}}\rangle_{2\mathrm{h}}$ by imposing $\langle \phi_0|c^{\dagger}_{i\uparrow}c^{\dagger}_{j\downarrow}| \Psi_{\text{inc}}\rangle_{2\mathrm{h}}= 0$ for all lattice sites, $i,j$. As in the single-hole case, these two components may be interpreted as arising from the average and fluctuation parts of the phase-twist operator
$e^{-im(\hat{\Omega}_i-\hat{\Omega}_j)}$ (see Appendix~\ref{app:fluctuation}).

Table.~\ref{tab:table2} lists the calculated VMC ground-state energy terms (in an $8\times 8$ system) associated with the two components in Eq.~(\ref{eq:2h_component}). The kinetic energy of the quasiparticle component $|\Psi_{\mathrm{Cooper}}\rangle_{2\mathrm{h}}$ is highly unfavorable ($\sim -0.012 J$) in comparison with that ($\sim -4.105J$) of the $|\Psi_{\text{inc}}\rangle_{2\mathrm{h}}$. But the latter is off-set by a higher superexchange energy cost ($\sim 3.354 J$) such that the total energy is lower than the former by $\sim -2 J$. The dominant kinetic (and total) energy contribution ($\sim -6.9 J$) to the two-hole ground state mainly comes from the cross matrix of $\hat{H}_t$ between the two components (cf. Table.~\ref{tab:table2}). 

\begin{figure*}[t]
\includegraphics[width=0.65\textwidth]{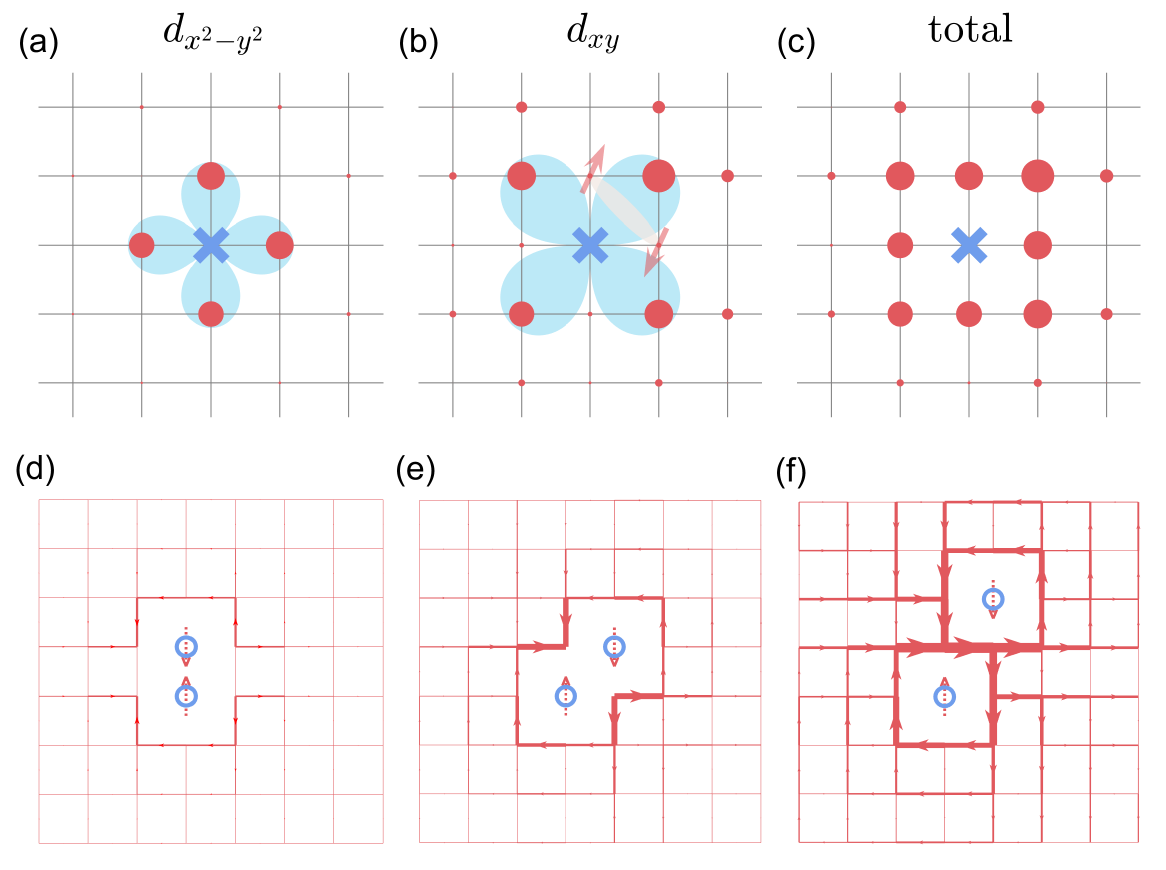}% Here is how to import EPS art
\caption{\label{fig:2h_ninjJs} 
(a–c) Expectation value of the hole-hole correlator $n^h_i n^h_j$ 
for the two-hole ground state: the size of each red circle indicates the weight of the correlator with one hole fixed at a given (blue cross) site $i$.  Panels show results for: (a) the quasiparticle (Cooper-pair) component $| \Psi_{\text{Cooper}}\rangle_{2\mathrm{h}}$, which corresponds to a $d_{x^2-y^2}$ pairing symmetry; (b) the incoherent component $| \Psi_{\text{inc}}\rangle_{2\mathrm{h}}$ which shows a $d_{xy}$ pairing symmetry with spin-singlet pairing created at the same sublattice sites (shown schematically with site $j$ located on the upper-right NNN site of the reference site $i$); and (c) the full variational wave function $|\Psi_{\mathrm{G}}\rangle_{2\mathrm{h}}$. 
Note that the slight deviation from perfect $C_4$ symmetry in hole-hole correlation pattern (a-c) arises from the off-centered placement of the reference site.
(d–f) Corresponding projected neutral spin-current pattern $(J_{ij}^s)_{kl,\text{proj}}$ between sites 
$i$ and $j$, with two holes projected at sites $k$ and $l$ created by $c_{k\uparrow}$ and $c_{l\downarrow}$ in the $m =- $ sector of Eq.~(\ref{eq:twohole}). The plotted quantity has been normalized by the hole-hole correlation $\langle n^h_k n^h_l \rangle$ to factor out the spatial dependence of the hole positions. Arrow thickness represents the relative current magnitude. All panels within each row are plotted on the same scale.} 
\end{figure*}

Namely, similar to the single-hole case, the two-hole ground state also forms a ``cat state'' with strong resonating between the Cooper and incoherent pairing state. Figures \ref{fig:2h_ninjJs}(a)-(c) illustrate the real-space hole-hole correlators, $\langle n^h_i n^h_j\rangle$ with $i$ fixed at the center of the system (blue cross), in Cooper and incoherent channels, as well as in the total ground state Eq.~(\ref{eq:2h_component}), respectively. It can be verified that $|\Psi_{\mathrm{Cooper}}\rangle_{2\mathrm{h}}$ contributes to a predominantly nearest-neighbor (NN) $d_{x^2-y^2}$ pairing in Fig.~\ref{fig:2h_ninjJs}(a) and $|\Psi_{\text{inc}}\rangle_{2\mathrm{h}}$ gives rise to the next-nearest-neighbor (NNN) $d_{xy}$ pairing in Fig.~\ref{fig:2h_ninjJs}(b), while the total density-density correlation in the non-degenerate ground state of $L_z=2\mod 4$ (with a $C_4$ rotational symmetry) is given in Fig.~\ref{fig:2h_ninjJs}(c). Similar hole-hole correlation structure has been reported in previous DMRG studies \cite{PhysRevB.55.6504,dpfl-12st}. Here, the two components are connected by the hopping term $\hat{H}_t$, with one hole hopping to its NN site, concomitant with a spin backflow such that a spin-singlet RVB pair at two opposite lattice sites in the half-filling ground-state $|\phi_0\rangle$ is transferred to a same-sublattice spin-singlet pairing, constituting an \emph{excitation} above $|\phi_0\rangle$ as indicated in Fig.~\ref{fig:2h_ninjJs}(b). The incoherent component refers to such a $d_{xy}$-wave Cooper pair of holes accompanied by the same-sublattice spin-singlet excitation, whose analysis will be given at the end of this subsection.

\begin{table}[t]
\caption{\label{tab:table2}%
The two-component contributions to the two-hole ground state energy on the $8 \times 8$ system by VMC method. 
Here the calculated $_{2\mathrm{h}}\langle \Psi_{\text{Cooper}}|\Psi_{\text{Cooper}}\rangle_{2\mathrm{h}}= 0.366$ and $_{2\mathrm{h}}\langle \Psi_{\text{inc}}|\Psi_{\text{inc}}\rangle_{2\mathrm{h}}= 0.634$.
}
\begin{ruledtabular}
\centering
\begin{tabular}{cccc}
\textrm{$\hat{O}=$}&
\textrm{$\hat{H}_{t\text{-}J}-E_{0}$}&
\textrm{$\hat{H}_t$}&
\textrm{$\hat{H}_J-E_{0}$}  \\  
\colrule
$_{2\mathrm{h}}\langle \Psi_{\text{Cooper}}|\hat{O}|\Psi_{\text{Cooper}}\rangle_{2\mathrm{h}}$ & $1.413$ & $-0.012$ & $1.426$ \\ 
$_{2\mathrm{h}}\langle \Psi_{\text{inc}}|\hat{O}|\Psi_{\text{inc}}\rangle_{2\mathrm{h}}$ & $-0.571$ & $-4.105$ & $3.534$\\ 
$ _{2\mathrm{h}}\langle \Psi_{\text{Cooper}}|\hat{O}|\Psi_{\text{inc}}\rangle_{2\mathrm{h}} +h.c.$ & $-6.967$  & $-6.949$  & $-0.019$ \\ 
$_{2\mathrm{h}}\langle \Psi_G|\hat{O}|\Psi_{G}\rangle_{2\mathrm{h}}$ & $-6.125$ & $-11.066$  & $4.941$ \\ 
\end{tabular}
\end{ruledtabular}
\end{table}

Figures \ref{fig:2h_ninjJs}(d)-(f) further show that when the two holes are further separated spatially in the ground-state wave function, the spin loop currents will reemerge for each $m=\pm $ in Eq. (\ref{eq:twohole}). Here the neutral spin current patterns $(J_{ij}^s)_{kl,\text{proj}}$ are calculated based on Eq. (\ref{eq:J_s}) with the spin-$\uparrow$ and spin-$\downarrow$ holes being projected at different locations $k$ and $l$ according to Eq. (\ref{eq:twohole}) at a given $m=-$ without loss of generality. 
Thus, the tight-binding of two holes is actually driven by the fusion of the incoherent components of the single holes to remove the loop currents. Due to the strong ``resonating'' between the quasiparticle and incoherent components, the Cooper pair of bare holes is formed merely as the consequence of such a fusion of the incoherent non-quasiparticle components of the holes. Therefore, the pairing mechanism here for two holes is fundamentally different from the usual BCS mechanism where the Landau quasiparticles are paired up to directly form a Cooper pair via exchanging a virtual bosonic mode, say, a spin-wave excitation above $|\phi_0\rangle$ in the spin-fluctuation theory \cite{PhysRevLett.69.961}. 

\begin{figure}[h]
\includegraphics[width=0.48\textwidth]{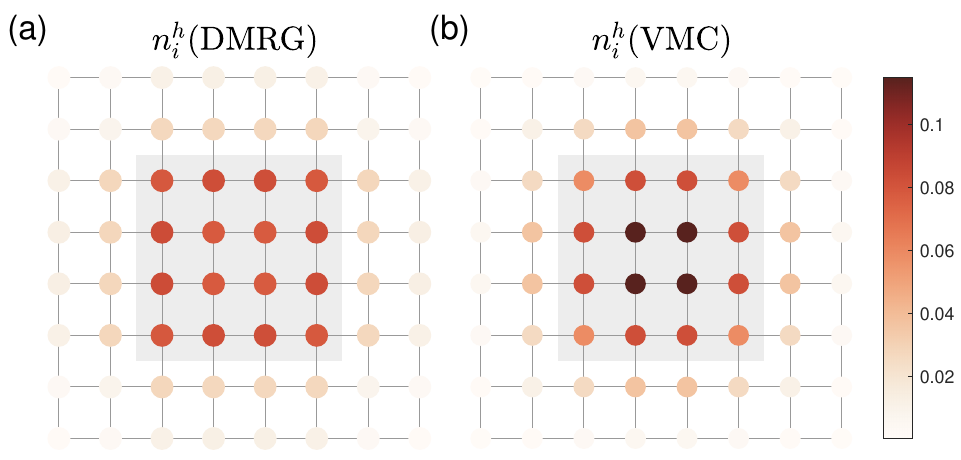}% Here is how to import EPS art
\caption{\label{fig:nhfusion} Hole density distribution of the two-hole ground state computed using (a) DMRG and (b) VMC.}
\end{figure}

The binding energy of the hole pair can be determined by calculating the energy difference between the first pair-broken excited state and the ground state of Eq.~(\ref{eq:twohole}), which is found \cite{PhysRevX.12.011062} by $E_{pair} \approx 2J$ with a scaling analysis showing the typical area occupied by the two holes is $3.8 a_0\times 3.8a_0$. Figure~\ref{fig:nhfusion} shows that the two-hole density distribution in the central $4 a_0\times 4a_0$ area of the $8\times 8$ system, calculated by DMRG and VMC method, respectively. (Note that in the VMC calculation, holes are confined to the central $8\times 8$ region of a $10\times 10$ lattice to suppress boundary localization and thereby reduce the energy penalty from broken superexchange bonds.)

\begin{figure}[t]
\includegraphics[width=0.35\textwidth]{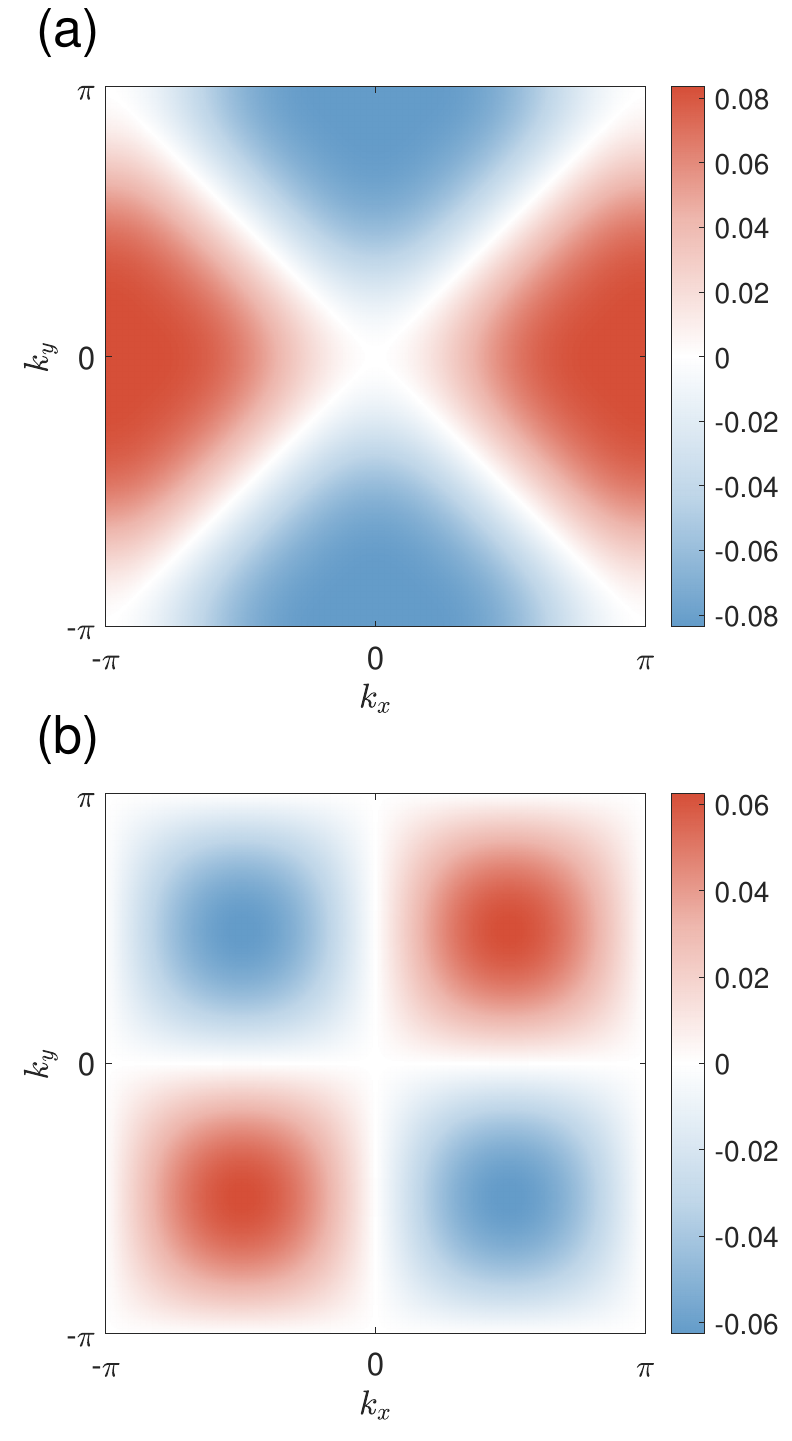}% Here is how to import EPS art
\caption{\label{fig:Deltak2h} (a) Cooper pair order parameter $\Delta_{\mathbf{k}}$ and (b) Imaginary part of the twisted pair order parameter $\mathrm{Im}(\tilde{\Delta}^{-}_{\mathbf{k}})$, revealing $d_{x^2-y^2}$ and $d_{xy}$ pairing symmetries, corresponding to $| \Psi_{\text{Cooper}}\rangle_{2\mathrm{h}}$ and $| \Psi_{\text{inc}}\rangle_{2\mathrm{h}}$, respectively. }
\end{figure}

Finally, in the momentum space, the Cooper-pair order parameter $\Delta_{\mathbf{k}}$ may be defined by 
\begin{equation}
\Delta_{\mathbf{k}}=_{\mathrm{2h}}\langle \Psi_{\mathrm{G}}|c_{\mathbf{k}\uparrow}c_{\mathbf{-k}\downarrow}-c_{\mathbf{k}\downarrow}c_{\mathbf{-k}\uparrow}|\phi_0\rangle~,
    \label{eq:Cooperpair}
\end{equation}
which indeed shows the $d_{x^2-y^2}$ pairing symmetry of the quasiparticles (Fig.~\ref{fig:Deltak2h}(a)) as evidenced by sign changes under $\frac{\pi}{2}$ rotation with the nodal lines along $k_x = \pm k_y$. This aligns with the $L_z = 2$ angular momentum of the two-hole wave function. This VMC calculation is carried out on a $12 \times 12$ lattice. 

However, for the incoherent component, due to its direct orthogonality with a Cooper pair channel, one has to explicitly take into account the aforementioned spin excitations in $|\Psi_{\text{inc}}\rangle_{2\mathrm{h}}$. The consequence is to have a $d_{xy}$-symmetry in a generalized pairing order parameter in ${\mathbf{k}}$-space as shown in Fig.~\ref{fig:Deltak2h}(b) with a sign reversal under $\frac{\pi}{2}$ rotation and the nodal lines along the $k_x=0$ and $k_y=0$ axes. 

Below we demonstrate the characterization of the pairing of the incoherent component. 
We may start by reexpressing the two-hole ground state \emph{Ansatz} in Eq.~(\ref{eq:twohole}) in the following form  
\begin{equation}
|\Psi_{\mathrm{G}}\rangle_{2\mathrm{h}}=\sum_{i,j,m=\pm 1}g'_{m}(i,j)c_{i\uparrow}c_{j\downarrow}e^{im\hat{A}^s_{ij}}|\phi_{0}\rangle, \label{eq:twohole_another}
\end{equation}
where $g_m'(i,j) \equiv g_m(i,j)e^{-im\theta_i(j)}e^{-im\phi^0_{ij}}$. Here, $\phi^0_{ij}$ and $\hat{A}^s_{ij}$ are defined as follows
\begin{equation}
    \phi^0_{ij}=\frac{1}{2} \sum_{l\neq i,j}\left[\theta_{i}(l)-\theta_{j}(l)\right],
    \label{eq:phiij}
\end{equation}
\begin{equation}
    \hat{A}^s_{ij}=\sum_{l\neq i,j}\left[\theta_{i}(l)-\theta_{j}(l)\right]S^z_l.
    \label{eq:Aijs}
\end{equation}
So it is the nontrivial phase fluctuations of $e^{+im\hat{A}^s_{ij}}$ that prevents $\Delta_{\mathbf{k}}$ from measuring the pairing symmetry in the incoherent sector $|\Psi_{\text{inc}}\rangle_{2\mathrm{h}}$. To probe the latter, one may introduce a ``twisted'' pairing order parameter: 
\begin{equation}
\tilde{\Delta}^{m}_{ij}=_{\mathrm{2h}}\langle \Psi_{\mathrm{inc}}|c_{i\uparrow}c_{j\downarrow}e^{+im\hat{A}^s_{ij}}-c_{i\downarrow}c_{j\uparrow}e^{+im\hat{A}^s_{ji}}|\phi_0\rangle, 
    \label{eq:twistedCooperpair}
\end{equation}
($m = \pm 1$) with the momentum-space representation $\tilde{\Delta}^{m}_{\mathbf{k}} = \frac{1}{N} \sum_{ij} e^{i\mathbf{k}\cdot(\mathbf{r}_i-\mathbf{r}_j)} \tilde{\Delta}^{m}_{ij}$. The calculated $\mathrm{Im}(\tilde{\Delta}^{-}_{\mathbf{k}})$ is presented in Fig.~\ref{fig:Deltak2h}(b) to indicate the $d_{xy}$ pairing symmetry. Here the focus on the imaginary part reflects the $d\pm id$-like feature of orbital wave functions $g'_{\mp}(i,j)$ in Eq. (\ref{eq:twohole_another}), which is also found in the previous honeycomb lattice study \cite{PhysRevB.110.155111}. After the summation over $m=\pm $, the incoherent component primarily hosts the real $d_{xy}$ pairing of two holes with the spin-singlet excitations, as weaker longer-distance pairing contributions are neglected. Note that the two-component $d$-wave pairing in the $c$-representation of the two-hole wave function is not contradictory with the $s$-wave pairing of the twisted quasiparticles $\tilde{c}_{i\sigma}$ mentioned in Ref.~\cite{PhysRevX.12.011062}, which is an equivalent description based on the more fundamental ``building block'' $\tilde{c}_{i\sigma}$.

\subsection{``Cat state'' revealed by spectral function}
\begin{figure*}[t]
\includegraphics[width=0.80\textwidth]{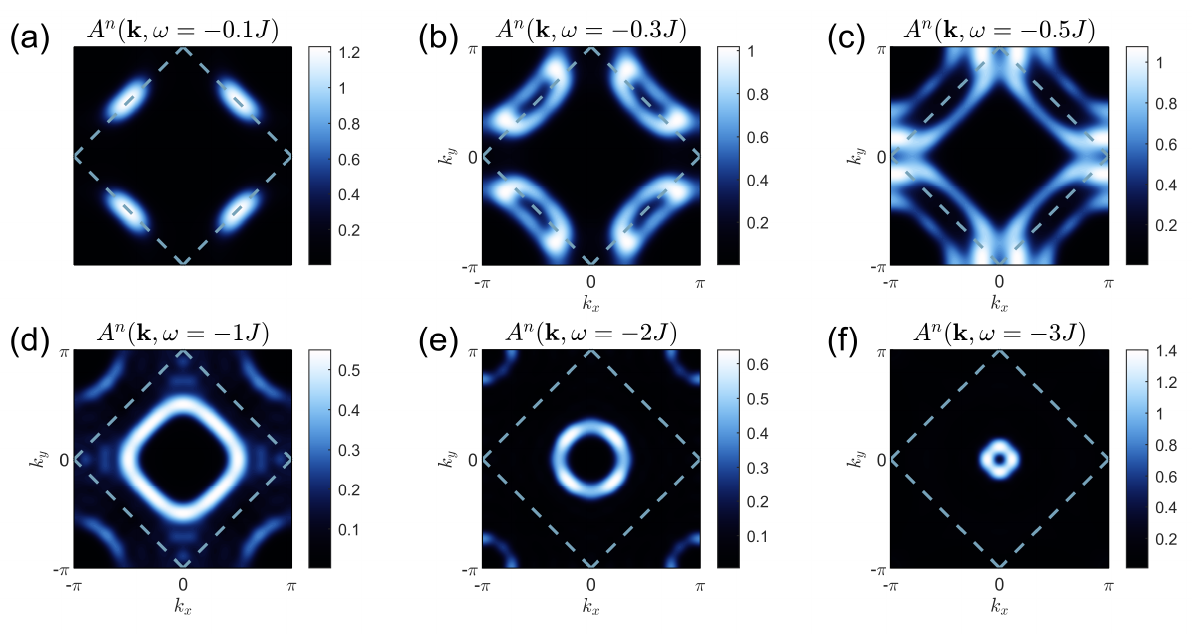}% Here is how to import EPS art
\caption{\label{fig:Akwnegcontour} 
Constant-energy contours of the single-particle spectral function $A^n(\mathbf{k},\omega)$ for a hole excited from the half-filled AFM ground state [see Fig.~\ref{fig:Akwneg} and Eq.~(\ref{eq:negbias})]. The contours illustrate the evolution of spectral weight in momentum space at fixed $\omega$. The blue dashed lines indicate the boundary of the magnetic Brillouin zone.} 

\end{figure*}

\begin{figure*}[t]
\includegraphics[width=0.80\textwidth]{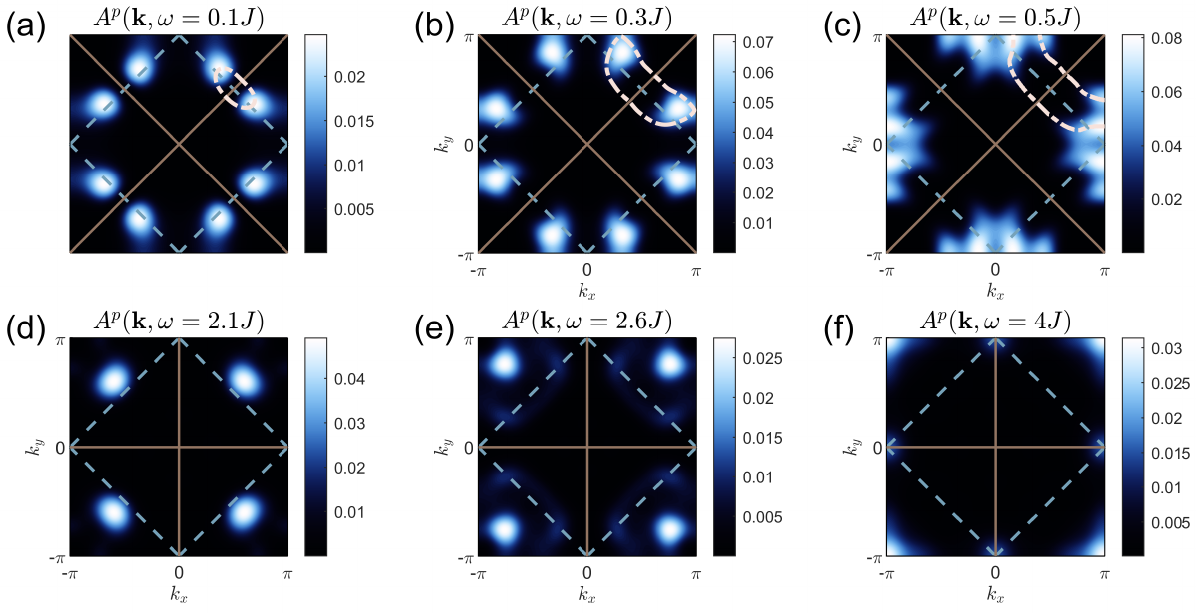}% Here is how to import EPS art
\caption{\label{fig:Akwpos} 
Single‑particle spectral function $A^p(\mathbf{k},\omega)$
for the two‑hole ground state [Eq.~(\ref{eq:Ap})]. (a-c) The dominant spectral weight peaks at the tips of the corresponding hole‑pocket contours (pink dashed curves, extracted from the single‑hole spectrum $A^n(\mathbf{k},\omega)$ in Fig.~\ref{fig:Akwnegcontour}).
These tips are separated by the nodal lines (brown) of $d_{x^2-y^2}$ symmetry. At higher energies ($\omega>2J$), spectral weight appears on the two sides of the nodal lines (brown) of $d_{xy}$ symmetry and shows no counterpart in the single‑hole spectrum, reflecting the contribution from the incoherent two‑hole pairing component (see text). Blue dashed lines mark the magnetic Brillouin‑zone boundary.}
\end{figure*}

In the previous section, it has been shown that for a single hole excited state $|\Psi_n\rangle_{\mathrm{1h}}$ created by injecting a hole into the half-filling ground state in an ARPES or STS probe, only its quasiparticle component can be detected by the spectral function $A^n(\mathbf{k},\omega)$ defined in Eq.~(\ref{eq:negbias}) at $\omega<0$, which is plotted in Fig.~\ref{fig:Akwneg}. The equal-$\omega$-contour of $A^n(\mathbf{k},\omega)$ is replotted in Fig.~\ref{fig:Akwnegcontour} at $t/J=3$. Four-Fermi-pocket-like contours centered at $(\pm \pi/2, \pm \pi/2)$ are shown in Figs.~\ref{fig:Akwnegcontour}(a)-(c) at low energies.  Figures ~\ref{fig:Akwnegcontour}(d)-(f) further show the high-energy trail of the quasiparticle component, which eventually disappears beyond the upper edge of the dispersion (cf. Fig.~\ref{fig:Akwneg}, where $t/J=2.5$ is slightly smaller). Here, the incoherent component of the single-hole state is absent in the spectral function due to the vanishing overlap in the structure factor $\left| _{\mathrm{1h}}\langle\Psi(n)|c_{\mathbf{k}\sigma}|\phi_{0}\rangle\right|^{2}$ in Eq.~(\ref{eq:negbias}), i.e., $\left| _{\mathrm{1h}}\langle\Psi_{\mathrm{inc}}(n)|c_{\mathbf{k}\sigma}|\phi_{0}\rangle\right|^{2} = 0$. 
Thus, a Landau-quasiparticle-like excitation may be wrongly concluded, simply because the crucial incoherent component of the ``cat state'' cannot be detected by the single-particle experiment as ``dark matter''  \cite{PhysRevB.111.104502}.

On the other hand, for the present two-hole ground state, it has been pointed out \cite{PhysRevB.111.104502} that the unconventional incoherent pairing component can be detected by the single-particle spectral function, which is defined by  
\begin{equation}\label{eq:Ap}
A^p(\mathbf{k},\omega)=-\mathrm{Im}\sum_n\frac{\left| _{\mathrm{1h}}\langle\Psi(n)|c_{\mathbf{k}\downarrow}^\dagger|\Psi_{\mathrm{G}}\rangle_{\mathrm{2h}}\right|^2}{\omega-[E_{\mathrm{1h}}(n)-E_{\mathrm{0}}^{\mathrm{2h}}+\mu_p]+i\eta},
\end{equation}
where $E_0^{\mathrm{2h}}$ denotes the two-hole ground state energy and $\mu_p$ is the chemical potential setting the excitation edge to zero. It corresponds to injecting an electron into the two-hole ground state to create a single-hole excitation state at frequency $\omega>0$, which can be measured by the inverse ARPES or the STS on the positive bias side. 

Here, the Cooper and non-Cooper parts in the two-hole ground state $|\Psi_{\mathrm{G}}\rangle_{\mathrm{2h}}$, can both contribute to a finite spectral weight $|_{\mathrm{1h}}\langle\Psi(n)|c_{\mathbf{k}}^\dagger|\Psi_{\mathrm{G}}\rangle_{\mathrm{2h}}|^2$ in Eq. (\ref{eq:Ap}). Namely, by injecting an electron into the two-hole ground state, a single-hole excitation that includes its incoherent component can be excited as characterized by
$A^p(\mathbf{k},\omega)$. The equal-$\omega$-contour of $A^p(\mathbf{k},\omega)$ at $\omega>0$ is shown in Fig.~\ref{fig:Akwpos},  which illustrates two distinct branches: low-energy branch $\omega<2J$ and high-energy branch $\omega>2J$, corresponding to the two-component structure of the two-hole ground state \cite{PhysRevB.111.104502}.  

For the low-energy branch,  both $A^n(\mathbf{k},\omega)$ at $\omega<0$ and $A^p(\mathbf{k},\omega)$ at $\omega>0$ show the symmetric energies for the same single hole excitation. However, their momentum contour structures exhibit drastic differences in Figs.~\ref{fig:Akwnegcontour}(a)-(c) and Figs.~\ref{fig:Akwpos}(a)-(c), respectively. Four-Fermi-pocket-like contours centered at 
$(\pm \pi/2, \pm \pi/2)$ are shown at $\omega<0$ in Figs.~\ref{fig:Akwnegcontour}(a)-(c). By contrast, at $\omega>0$ in Figs.~\ref{fig:Akwpos}(a)-(c), $d_{x^2-y^2}$ nodal lines cut through the hole pockets such that densities of the state vanish in the nodal regions, which are now redistributed and mainly concentrate on the ``tips'' of the pockets, consistent with the $d_{x^2-y^2}$-wave symmetry of $|\Psi_{\mathrm{Cooper}}\rangle_{2\mathrm{h}}$ in the two-hole ground state discussed before.  

At higher energies ($\omega = 2J - 4J$), $A^p(\mathbf{k},\omega)$ in Figs.~\ref{fig:Akwpos}(d)-(f) further shows that spectral weight maxima now concentrate along the $k_x = \pm k_y $ lines, which systematically shift from $(\pm \pi/2, \pm \pi/2)$ at $\omega=2J$ toward the $(\pi,\pi)$ point at $\omega=4J$, with suppressed intensities along $k_x=0$ and $k_y=0$ consistent with $d_{xy}$ pairing symmetry in $|\Psi_{\text{inc}}\rangle_{2\mathrm{h}}$. This is in contrast to $A^n(\mathbf{k},\omega)$ at $\omega<0$ in Figs.~\ref{fig:Akwnegcontour}(d)-(f), which only show the high-energy trail of the quasiparticle component as noted above.

\section{Summary and discussion}\label{sec:6}

Through a systematic VMC study, we reveal novel properties of doped holes in Mott-AFM insulators. Our approach is based on wave functions constructed from an accurate half-filled ground state of the Heisenberg model \cite{Liang1988}, 
augmented by a duality transformation that explicitly tracks the singular phase-string sign structure of the 
$t$-$J$ model upon doping. The VMC results are validated against numerical benchmarks. 

One of our key findings is the identification of the {\bf single doped hole as a non-Landau quasiparticle}, characterized by:

\begin{itemize}
    \item {\bf A two-component ``cat state'' structure}: A quantum superposition of a bare hole and an incoherent part involving spin and charge loop currents.
\end{itemize}

\begin{itemize}
    \item {\bf Resonant energy gain}: The system's kinetic energy is predominantly gained through resonance between these two components, making the state non-perturbative in nature.
\end{itemize}

\begin{itemize}
    \item {\bf Distinguishing signatures}: This state exhibits an emergent angular momentum $L_z=\pm 1$, a loop magnetic moment $m_{\mathrm{loop}}\simeq 0.1 \mu_B$, and a $4a_0\times 4a_0$ spatial modulation. These features explicitly demonstrate the breakdown of a one-to-one correspondence with a conventional Bloch wave, confirming its non-Landau nature.

\end{itemize}

Another key finding is the instability of single holes toward pair formation, leading to {\bf a tightly bound, non-Cooper-like pair}. This two-hole ground state is characterized by:
\begin{itemize}
    \item {\bf A two-component ``cat state'' structure}: A quantum superposition of a $d_{x^2-y^2}$-wave Cooper pair of bare holes at NN sites, and an incoherent component involving $d_{xy}$-like pairing at NNN sites together with spin-singlet excitations.  

\end{itemize}

\begin{itemize}
    \item {\bf Resonant binding mechanism}: The ground‑state energy is optimized by a strong resonance between these two components. Here, the incoherent component facilitates the fusion of the two holes, while the Cooper pairing arises primarily from resonant coupling between the sectors, mediated by the hopping term.    
\end{itemize}

\begin{itemize}
    \item {\bf Distinguishing signatures}: The bound pair carries an emergent angular momentum $L_z=2$ and exhibits a compact spatial profile, typically confined within a $\sim 4a_0\times 4a_0$ region.
\end{itemize}

Here, the single doped hole can be approximately described by an effective Wannier-like wave function spanning a $4a_0\times 4a_0$ plaquette, rather than being associated with a single lattice site. Similarly, the two-hole bound state extends over a characteristic region of comparable size in large systems. Both of these emergent length scales remain finite and are significantly smaller than the divergent spin-spin correlation length $\xi$ of the antiferromagnetic background $|\phi_{0}\rangle$. As a result, long-wavelength spin correlations in $|\phi_{0}\rangle$ do not directly govern the formation of the non-Landau quasiparticle or the non-Cooper-pair ``cat state''. Instead, the behavior of the doped holes is dominated by strong entanglement with the local spin environment, mediated by the transverse dipolar phase factor $e^{-im\left(\hat{\Omega}_{i}-\hat{\Omega}_{k}\right)}$. In this description, the hole and its bound antimeron [Eq. (\ref{eq:singlehole})]—or the two holes forming a pair [Eq. (\ref{eq:twohole})]—reside at positions $i$ and $k$, which precisely control the singular phase-string effect arising from hole motion in the $t$–$J$ model. Indeed, suppressing the phase-string sign structure—as in the $\sigma\cdot t$-$J$ model, where the factor $e^{-im\left(\hat{\Omega}_{i}-\hat{\Omega}_{k}\right)}$ is absent—restores conventional Landau quasiparticle behavior, as demonstrated in Appendix ~\ref{app:longitudinal}. In summary, the non-Landau nature of the single hole and the unconventional pairing mechanism for two holes both stem from quantum interference effects inherent to the phase-string sign structure of the $t$–$J$ model.

Experimental detection of non-Landau quasiparticles or non-Cooper pairs in lightly doped Mott insulators remains a significant challenge. For example, while a single-hole state can be introduced into the ground state $|\phi_{0}\rangle$ by electron removal in ARPES or STS, the measured spectral function $A^n({\mathbf k},\omega)$ (Figs. \ref{fig:Akwneg} and \ref{fig:Akwnegcontour}) captures only the bare-hole component. The incoherent loop-current portion remains hidden due to wave function orthogonality. This is illustrated by the real-space wave function in Fig.~\ref{fig:pwave}: a local STS probe at low energies would exhibit finite quasiparticle spectral weight only at the red-dot sites, while vanishing at the $p$-orbital-like (blue) incoherent sites. As a result, the single-hole non-Landau state could be misinterpreted experimentally as a conventional—but translation-symmetry-breaking—quasiparticle, with its essential incoherent part acting as undetected “dark matter''. A related effect was shown previously \cite{PhysRevB.111.104502}: a hole trapped near an off-site impurity displays a four-lobe clover pattern in STS \cite{ye2023visualizingzhangricesingletmolecular} at low energy due to a similar orthogonality catastrophe, even though its actual density is distributed much more uniformly around the impurity and can only be recovered by the STS at high energies.

For the two-hole pairing state, the Cooper-pair component is reflected in the spectral function $A^p({\mathbf k},\omega)$, accessible via inverse ARPES or positive-bias STS at low energy. In contrast, signatures of the incoherent pairing component appear in the \emph{high-energy branch} of $A^p({\mathbf k},\omega)$ [Figs.~\ref{fig:Akwpos}(d)-(f)]. Here, an injected electron annihilates one hole from a pair in the ground state, generating single-hole excited states that contribute finite spectral weight to $A^p({\mathbf k},\omega)$. This signal can be observed in STS as a high-energy peak above the pair-breaking energy $E_{\mathrm{pair}}$ \cite{PhysRevB.111.104502}, alongside the low-energy “coherent'' peak linked to nodal-quasiparticle-like excitations [Figs.~\ref{fig:Akwpos}(a)-(c)]. In particular, sharp “banana-tip'' features in the density of states—enhanced by the $d$-wave pairing symmetry—are expected to produce pronounced quasiparticle interference (QPI) patterns within each $4\times 4 $ hole-pair block in real space (cf. Fig.~\ref{fig:nhfusion}). Once these $4\times 4 $ building blocks pack into larger quasiperiodic arrangements, strong QPI signals may emerge at wavevectors that resonate between these spectral peaks along the spatial confined direction. Such emergent quantum interference effects beyond the QPI of conventional Landau/Bogoliubov quasiparticles may provide an alternative explanation for the local STS observation \cite{Ye2023,Li2023} of novel internal checkerboard/stripe structures associated with two-hole plaquettes in the hole puddles of the underdoped cuprate.

The proposed picture further yields several testable consequences in the lightly doped AFM phase. First, the minimal $2\times 2$ loop-current pattern imparts an anomalous magnetic moment $m_{\mathrm{loop}}\sim 0.1 \mu_B$ to the single-hole composite. Polarized by a perpendicular field, this moment allows a temperature gradient to induce a thermal Hall signal upon doping \cite{Grissonnanche2019,hu2025highresolutionmeasurementsthermalconductivity}. 
Second, a Rashba spin-orbit coupling can lock the hole’s orbital angular momentum $L_z = \pm 1$
to its spin $S^z = \pm 1/2$, defining a total $J_z = L_z + S^z$ \cite{PhysRevB.105.075136}. The resulting spin textures provide a distinct signature for spin-resolved ARPES \cite{Gotlieb2018,Luo2024,Iwasawa2023}.

Finally, we briefly outline a possible route to superconductivity at finite doping, starting from the present two‑hole ground state as a minimal pairing building block. Eq.~(\ref{eq:twohole}) can be recast as $|\Psi_{\mathrm{G}}\rangle_{\mathrm{2h}}=\hat{\mathfrak{D}}|\phi_0\rangle $ with  
$$\hat{\mathfrak{D}} = \sum_{i,j,m=\pm 1} g_m(i,j)c_{i\uparrow}c_{j\downarrow}e^{-im\left(\hat{\Omega}_{i}-\hat{\Omega}_{j}\right)}. $$ Within the AFM phase, one may conjecture a finite‑doping state of the form 
\begin{equation}
|\Psi_{\mathrm{G}}\rangle_{\mathrm{AFM}} = \left(\hat{\mathfrak{D}}\right)^{N_h/2}|\phi_0\rangle \label{eq:finitedopingnew}
\end{equation}
where $N_h$ is the total number of holes, implying a condensation of $\langle \hat{\mathfrak{D}} \rangle\neq 0$ in the thermodynamic limit. However, because each hole pair described by $\hat{\mathfrak{D}}$ forms a ``cat state''—a strong resonance between a mobile Cooper pair and an incoherent component confined to a region of size $\sim 4a_0\times 4a_0$—such a finite‑doping state may actually be spatially confined or localized as an insulator, a point that requires further examination \cite{PhysRevX.12.011062}. 

One may further conjecture that as the density of hole pairs increases at low doping, these localized hole‑rich puddles begin to percolate, forming domains of local phase coherence, as hinted by STS experiments \cite{Ye2023}. At sufficiently high doping, this process may culminate in a uniform superconducting ground state \cite{Weng_2011}:
\begin{equation}
|\Psi_{\mathrm{G}}\rangle = \left({\hat{\mathcal{D}}}\right)^{N_h/2}|\mathrm{RVB}\rangle \label{eq:finitedoping}~,
\end{equation}
where now $\hat{\mathcal{D}} = \sum_{i,j} g(i,j)\tilde{c}_{i\uparrow}\tilde{c}_{j\downarrow}$ creates pairs of twisted holes $\tilde{c}_{i\sigma } = c_{i\sigma} e^{-i\hat{\Omega}_i}$, and the background spin state $|\mathrm{RVB}\rangle$ becomes an emerging short‑range RVB liquid as a self-consistent result of the condensation $\langle \hat{\mathcal{D}} \rangle$ \cite{Weng_2011}.

The transition from Eq.~(\ref{eq:finitedopingnew}) to Eq.~(\ref{eq:finitedoping}) can be understood by rewriting $$\hat{\mathfrak{D}} = \sum_{i,j} \tilde{c}_{i\uparrow}\tilde{c}_{j\downarrow}\left(g_{+}(i,j)e^{ i2\hat{\Omega}_{j}}+g_{-}(i,j)e^{i2\hat{\Omega}_{i}}\right)$$ and noting that \footnote{In the previous approach \cite{Weng_2011}, $\tilde{c}_{i\sigma } = c_{i\sigma} e^{-i\hat{\Omega}_i}$ is specifically taken. It is noted that one may also define the twisted hole operator $\tilde{c}_{i\sigma } = c_{i\sigma} e^{i\hat{\Omega}_i}$ by changing the sign in front of $\hat{\Omega}_i$. Correspondingly, one has $$\hat{\mathfrak{D}} = \sum_{i,j} \tilde{c}_{i\uparrow}\tilde{c}_{j\downarrow}\left(g_{-}(i,j)e^{-i2\hat{\Omega}_{j}}+g_{+}(i,j)e^{-i2\hat{\Omega}_{i}}\right)$$ such that a mean-field solution of $|\mathrm{RVB}\rangle $ associated with the double‑antimeron field $e^{-i2\hat{\Omega}_{v}}$ with an opposite chirality can be obtained. Two ground states have equal energy without time-reversal symmetry breaking \cite{Weng_2011}. A possible topological order in the superconducting phase due to such double choices of the chirality will be discussed elsewhere. }
$$\left(g_{+}(i,j)e^{i2\hat{\Omega}_{j}}+g_{-}(i,j)e^{i2\hat{\Omega}_{i}}\right)|\phi_0\rangle \rightarrow g(i,j)|\mathrm{RVB}\rangle ,$$
where $g(i,j)$ is an $s$-wave pair amplitude \cite{Ma2014,PhysRevX.12.011062}. 
Physically, the double‑antimeron field $e^{i2\hat{\Omega}_{v}}$—originally bound to the twisted‑hole pair with $v\in (i,j)$—becomes deconfined and smears throughout the spin background, thereby converting the long‑range AFM state $|\phi_0\rangle$ into the short‑range RVB state $|\mathrm{RVB}\rangle $ \cite{Weng_2011,Ma2014}.  Self-consistently, the proliferation and condensation of the deconfined twisted‑hole pairs with $\langle \hat{\mathcal{D}} \rangle\neq 0$, on top of the RVB background, give rise to the superconducting and lower pseudogap phases proposed in Ref. \cite{Weng_2011}. The nature of the highly unconventional transition between the AFM and emergent RVB phases remains an important open question, reserved for future investigation.

\begin{acknowledgments}

Useful discussions with H.-K. Zhang, J.-X. Zhang, S. Chen, Y. Qi, Y.Y. Wang, H.H. Wen, X.J. Zhou, Z.X. Shen, D.N. Sheng, J. Zaanen, R.B. Laughlin, T. Devereaux, C. Varma, and P.A. Lee are acknowledged. This work is supported by NSF of China (Grant No. 12347107) and MOST of China (Grant No. 2021YFA1402101).

\end{acknowledgments}

\appendix

\section{\label{app:secA} Loop current patterns in the single-hole ground state as calculated by DMRG and VMC}
\begin{figure*}[t]
\includegraphics[width=0.5\textwidth]{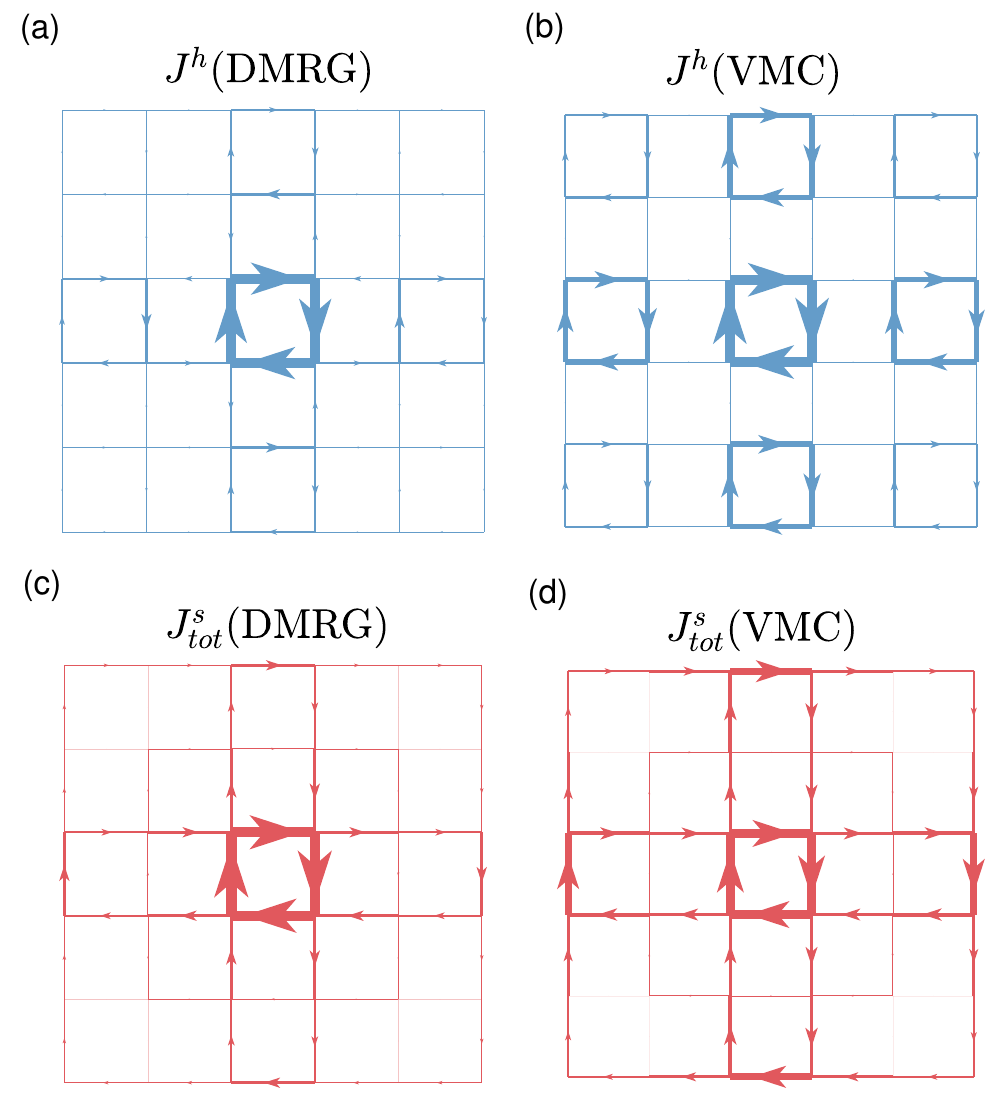}% Here is how to import EPS art
\caption{\label{fig:currentlx6} 
Hole and spin currents for the degenerate one-hole ground state ($L_z = 1$ and $S^z =-\frac{1}{2}$). Calculated hole current $J^h$ (a, b) and total spin current $J^s_{\mathrm{tot}}$ (c, d) on a $6\times 6$ lattice. Results from DMRG (a, c) and VMC (b, d) are shown, respectively. Arrow thickness indicates relative current strength; hole and spin currents are displayed on separate scales.}
\end{figure*}

We perform DMRG calculations for the single‑hole‑doped 
$t$-$J$ model using the ITensor library \cite{10.21468/SciPostPhysCodeb.4-r0.3}. For the single‑hole case the bond dimension is kept up to 
$D = 5000$, yielding a maximum truncation error 
$\epsilon \sim 5.5\times 10^{-6}$; for the two‑hole case we maintain 
$\epsilon \sim 1.6\times 10^{-5}$. The U(1) spin symmetry is enforced by fixing the total 
$S^z = -\frac{1}{2}$ for the single-hole case (without loss of generality).
For the degenerate single-hole ground states, we first compute one ground state $|\Psi_{1}\rangle$; a second ground state $|\Psi_{2}\rangle$, orthogonal to  $|\Psi_{1}\rangle$, is then obtained by adding an energy penalty that suppresses any overlap with $|\Psi_{1}\rangle$. The two resulting real wave functions can be combined into the complex states
\begin{equation}
|\Psi_{\pm}\rangle_{\mathrm{1h}}=\frac{1}{\sqrt{2}} (|\Psi_{1}\rangle\pm i|\Psi_{2}\rangle)~,
\end{equation}
which are eigenstates with angular momenta $L_z = \pm 1$ under $C_4$-rotation symmetry.

Figure~\ref{fig:currentlx6} displays the hole‑current and spin‑current patterns for the 
$L_z = 1$ ground state on a $6\times 6$ lattice, as obtained from DMRG and VMC. In contrast to the $8 \times 8$ system (Fig.~\ref{fig:currenttotal}), where the four strongest current loops are symmetrically arranged around the center, the $6\times 6$ pattern exhibits a single dominant $2 \times 2$ current loop located at the center of the lattice.

\section{\label{app:longitudinal}Variational results based on the quasiparticle picture}

In this appendix, we demonstrate that the conventional Landau quasiparticle picture fails to capture key nontrivial properties of the single-hole ground state. For comparison with the phase-string-based Ansatz in Eq.~(\ref{eq:singlehole}), we first consider a simple Bloch-wave trial state:
\begin{equation}
|\Psi_{\text{Bloch}}\rangle_{1\mathrm{h}}=\sum_{i}\varphi_{\mathrm{B}}(i)c_{i\uparrow}|\phi_{0}\rangle. \label{eq:Bloch}
\end{equation}
To incorporate the renormalization of the holon’s effective mass due to holon‑magnon scattering (the “longitudinal spin‑polaron'' effect), we further improve the Bloch‑wave Ansatz via a first‑order Lanczos step, leading to the following variational state:
\begin{equation}
\begin{split}
|\Psi_{\text{Lanczos}}\rangle_{1\mathrm{h}} = &\sum_{i}{\Big(\varphi_{1}(i)c_{i\uparrow}+\sum_{j\in NN(i)}(\varphi_{2}(i,j)n_{i\uparrow}c_{j\uparrow}\Big. } \\
&{\Big. +\varphi_{3}(i,j)S^{-}_{i}c_{j\downarrow})\Big)}|\phi_0\rangle.
\label{eq:lanczos}
\end{split}
\end{equation}
Table~\ref{tab:tablecomparison} summarizes the total energy, kinetic and superexchange contributions, and the angular momentum $L_z$ for each variational state. The plain Bloch‑wave Ansatz yields a kinetic energy of approximately $-2.158$, which is notably poor. Incorporating the longitudinal spin‑polaron effect significantly improves the kinetic energy to about $-6.411$; however, the resulting ground state is unique and carries $L_z = 2$, in contrast to the doubly degenerate $L_z = \pm 1$ states identified by DMRG. These results confirm that the nontrivial $L_z = \pm 1$ degeneracy—together with the accompanying charge‑ and spin‑current patterns—can only be captured by including the mutual statistics between holon and spinon (i.e., the phase‑string sign structure), as implemented in our Ansatz in Eq.~(\ref{eq:singlehole}).

\begin{table}[h]
\caption{\label{tab:tablecomparison}%
Comparison of single‑hole ground‑state energies (in units of $J$) and angular momentum 
$L_z$ on an $8 \times 8$ lattice for the $t–J$ model, obtained from VMC and DMRG. $E_{\text{tot}}$, $E_t$, and $E_J$ denote the total, kinetic, and superexchange energies, respectively.
}
\begin{ruledtabular}
\begin{tabular}{ccccc}
&
\textrm{$E_{\text{tot}}$}&
\textrm{$E_t$}&
\textrm{$E_J$}&
\textrm{$L_z$}\\ 
\colrule
$|\Psi_{\text{Bloch}}\rangle_{1\mathrm{h}}$  &  -67.98 & -2.16 & -65.82 & 0\\
$|\Psi_{\text{Lanczos}}\rangle_{1\mathrm{h}}$  &  -71.42 & -6.41 & -65.00 & 2\\
DMRG &  -72.98 & -8.64 & -64.34 & $\pm 1$\\
\end{tabular}
\end{ruledtabular}
\end{table}

To further demonstrate the nontrivial effects arising from the phase string, we “turn off'' this effect by modifying the hopping term $\hat{H}_t$ to $\hat{H}_{\sigma \cdot t}$, where a spin‑dependent sign $\sigma$
is attached to the hopping process:
\begin{equation}
	\hat{H}_{\sigma \cdot t} = -t\sum_{\langle ij \rangle,\sigma}\sigma c^{\dagger}_{i\sigma}c_{j\sigma} +h.c., \label{eq:sigmatJ}
\end{equation}
while the superexchange term $\hat{H}_J$ remains unchanged. This $\sigma\cdot t -J$ model shares the same Marshall‑sign structure \cite{Marshall1955-gi} with the half‑filled Heisenberg Hamiltonian and exhibits no phase frustration in the single‑hole hopping process \cite{Weng1997}.

Table.~\ref{tab:tablesigmatJ} presents the ground‑state energy and angular momentum calculated for this modified model using $|\Psi_{\text{Bloch}}\rangle_{1\mathrm{h}}$, $|\Psi_{\text{Lanczos}}\rangle_{1\mathrm{h}}$, and DMRG. DMRG confirms that the ground state is unique with $L_z = 2$ and exhibits no current patterns, reinforcing the connection between nontrivial current patterns and the phase‑string effect. The Bloch‑wave Ansatz correctly reproduces this quantum number, and including the longitudinal spin‑polaron effect in $|\Psi_{\text{Lanczos}}\rangle_{1\mathrm{h}}$ further improves the kinetic energy (to approximately $-10.398$), bringing it close to the DMRG result ($-10.918$). The spectral weight $Z_{\mathbf{k}}$ calculated by both VMC and DMRG shows a sharp peak at $(\pi,\pi)$ for total $S^z = -\frac{1}{2}$ and at $(0,0)$ for total $S^z = \frac{1}{2}$, indicating that the single‑hole ground state in the $\sigma\cdot t$-$J$ model exhibits conventional Landau‑quasiparticle behavior once the singular $(\pm 1)$ phase‑string signs in the hopping process are eliminated.

\begin{table}[h]
\caption{\label{tab:tablesigmatJ}%
Comparison of single-hole ground-state energies (in units of $J$) and angular momentum $L_z$ for the $\sigma\cdot t$-$J$ model on an $8 \times 8$ lattice with total $S^{z} = -1/2$, calculated using VMC and DMRG. $E_{\text{tot}}$, $E_t$, and $E_J$ denote the total, kinetic, and superexchange energies, respectively.}
\begin{ruledtabular}
\begin{tabular}{ccccc}
&
\textrm{$E_{\text{tot}}$}&
\textrm{$E_{t}$}&
\textrm{$E_J$}&

\textrm{$L_z$}\\ 
\colrule
$|\Psi_{\text{Bloch}}\rangle_{1\mathrm{h}}$   & -73.49 &  -8.17 & -65.32  & 2\\
$|\Psi_{\text{Lanczos}}\rangle_{1\mathrm{h}}$  & -75.58 &  -10.40 & -65.18  & 2\\
DMRG & -76.08 & -10.92  & -65.17  & 2\\
\end{tabular}
\end{ruledtabular}
\end{table}

\begin{figure*}[t]
\centering
\includegraphics[width=0.70\textwidth]{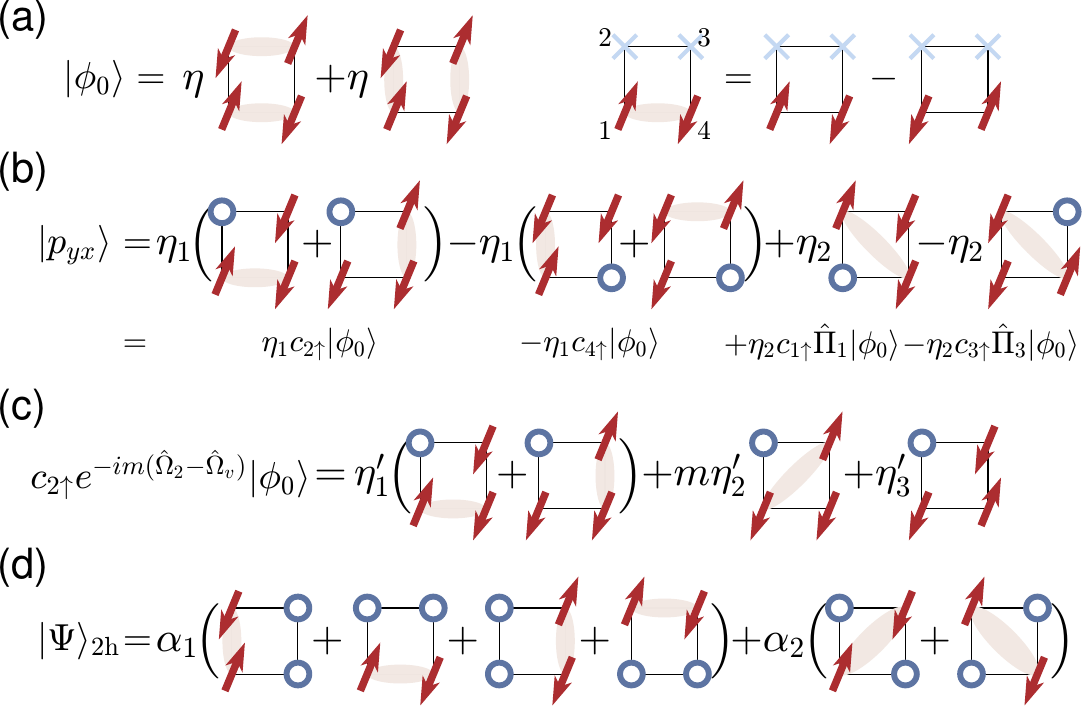}
\caption{\label{fig:EDsolution}(a) The half-filled ground state $|\phi_0\rangle$. $\eta = 0.289$. Bold lines indicate spin-singlet pairs $|\uparrow_i \downarrow_j\rangle -|\downarrow_i\uparrow_j\rangle $, and the light-blue crosses represent the spin/hole. The states are expressed under Ising basis: $c^{\dagger}_{l_1, \sigma_{l_1}}c^{\dagger}_{l_2, \sigma_{l_2}}\dots c^{\dagger}_{l_{N_e}, \sigma_{l_{N_e}}} |0\rangle$ where $l_1 < l_2 < \dots< l_{N_e}$ and $N_e$ is the total particle number. (b) Configuration of the $|p_{yx}\rangle$ ground state in the intermediate range ($0.5<t/J<3.8$). $\eta_1 = 0.214$ and $\eta_2 = 0.336$ when $t/J = 3$. (c) Representation of $c_{i\uparrow}e^{-im(\hat\Omega_i-\hat\Omega_v)}|\phi_0\rangle$ for $i = 2$ ($m =\pm1$), with $\eta_1' = 0.204$, $\eta_2' = -0.204i$, $\eta_3' = -0.149$. (d) The two-hole doped ground state $|\Psi\rangle_{\mathrm{2h}}$ in the $2\times 2$ system, with $\alpha_1 = 0.257$ and $\alpha_2 = 0.343$ when $t/J = 3$. }
\end{figure*}

\section{\label{app:fluctuation}Another perspective on the two components in the ``cat state''}

In the main text, we decompose the single-hole and two-hole wave function into two orthogonal components. Here, we point out that the quasiparticle (or Cooper-pair) and incoherent components in both wave functions can be naturally interpreted as the average part $\langle e^{-im\left(\hat{\Omega}_{i}-\hat{\Omega}_{k}\right)}\rangle_{\text{avg}}$ and fluctuation part $: e^{-im\left(\hat{\Omega}_{i}-\hat{\Omega}_{k}\right)}: \equiv  e^{-im\left(\hat{\Omega}_{i}-\hat{\Omega}_{k}\right)} - \langle e^{-im\left(\hat{\Omega}_{i}-\hat{\Omega}_{k}\right)}\rangle_{\text{avg}}$ of the phase-twist operator respectively.

In the single-hole case, we can rewrite the single-hole wave function as:
\begin{equation}
\begin{split}
    |\Psi\rangle_{1\mathrm{h}} &= \sum_{m=\pm 1,v}\varphi_m(i,v)c_{i\uparrow}\Bigl(\langle e^{-im\left(\hat{\Omega}_{i}-\hat{\Omega}_{v}\right)}\rangle_{\text{avg}}\\
    &\,\,\,\,\,\,\,\,\,\,\,\,\,\,\,\,\,\,\,\,\,\,\,\,\,\, +: e^{-im\left(\hat{\Omega}_{i}-\hat{\Omega}_{v}\right)}:\Bigr)|\phi_0\rangle \\
     &\equiv |\Psi_{\text{qp}}\rangle_{1\mathrm{h}} + |\Psi_{\text{inc}}\rangle_{1\mathrm{h}}, \label{eq:1h_decomp}
\end{split}
\end{equation}
where the definition of the average part is:
\begin{equation}
    \langle e^{-im\left(\hat{\Omega}_{i}-\hat{\Omega}_{v}\right)}\rangle_{\text{avg}} \equiv \frac{\langle \phi_0|n_{i\uparrow} e^{-im\left(\hat{\Omega}_{i}-\hat{\Omega}_{v}\right)}|\phi_0\rangle}{\langle \phi_0
|n_{i\uparrow}|\phi_0 \rangle}
\end{equation}
Here, the $n_{i\uparrow}$ operator in the numerator restricts the average in the Hilbert subspace with a spin-$\uparrow$ electron at site $i$ and $\langle \phi_0
|n_{i\uparrow}|\phi_0 \rangle = 0.5$ in the denominator is the normalization factor.

Therefore, the coefficient $\varphi(i)$ in quasiparticle component  $|\Psi_{\text{qp}}\rangle_{1\mathrm{h}}=\sum_i\varphi(i)c_{i\sigma} |\phi_0\rangle$ can be obtained by:
\begin{equation}
    \varphi(i) = \sum_{m=\pm 1,v}\varphi_m(i,v)\langle e^{-im\left(\hat{\Omega}_{i}-\hat{\Omega}_{v}\right)}\rangle_{\text{avg}}
\end{equation}
which is equivalent to the expression given in the main text $\varphi(i) = 2\langle \phi_0 | c_{i\sigma}^{\dagger} |\Psi_{\mathrm{G}}\rangle_{1\mathrm{h}}$.

Similarly, we can also rewrite two-hole wave function into two components:
\begin{equation}
\begin{split}
    |\Psi\rangle_{2\mathrm{h}} &= \sum_{i,j,m=\pm 1}g_m(i,j)c_{i\uparrow}c_{j\downarrow}\Bigl( \langle e^{-im\left(\hat{\Omega}_{i}-\hat{\Omega}_{j}\right)}\rangle_{\text{avg}}\\
    &\,\,\,\,\,\,\,\,\,\,\,\,\,\,\,\,\,\,\,\,\,\,\,\,\,\, +: e^{-im\left(\hat{\Omega}_{i}-\hat{\Omega}_{j}\right)}:\Bigr)|\phi_0\rangle \\
    &\equiv |\tilde{\Psi}_{\text{Cooper}}\rangle_{2\mathrm{h}} + |\tilde{\Psi}_{\text{inc}}\rangle_{2\mathrm{h}}, \label{eq:2h_decomp}
\end{split}
\end{equation}
where
\begin{equation}
    \langle e^{-im\left(\hat{\Omega}_{i}-\hat{\Omega}_{j}\right)}\rangle_{\text{avg}} \equiv \frac{\langle \phi_0|n_{i\uparrow}n_{j\downarrow} e^{-im\left(\hat{\Omega}_{i}-\hat{\Omega}_{j}\right)}|\phi_0\rangle}{\langle \phi_0
|n_{i\uparrow}n_{j\downarrow}|\phi_0 \rangle}.
\end{equation}
Although the decomposition in Eq.~(\ref{eq:2h_decomp}) does not strictly guarantee orthogonality between the two components, their overlap is numerically negligible, with $_{2\mathrm{h}}\langle \tilde{\Psi}_{\text{Cooper}}|\tilde{\Psi}_{\text{inc}}\rangle_{2\mathrm{h}} = -0.004$. Moreover, the resulting components show close agreement with those obtained from the strict orthogonal decomposition in Eq.~(\ref{eq:2h_component}): the overlaps $_{2\mathrm{h}}\langle \tilde{\Psi}_{\text{Cooper}}|\Psi_{\text{Cooper}}\rangle_{2\mathrm{h}}= 0.370$ and $_{2\mathrm{h}}\langle \tilde{\Psi}_{\text{inc}}|\Psi_{\text{inc}}\rangle_{2\mathrm{h}}= 0.634$ are nearly identical to the corresponding weights reported in Table~\ref{tab:table2}. This close quantitative consistency indicates that the Cooper-pair and incoherent components in Eq.~(\ref{eq:2h_component}) can equivalently be interpreted as the average and fluctuation parts of the phase-twist operator.

\section{\label{app:ED}ED analysis on the plaquette state}

\begin{figure}[t]
\includegraphics[width=0.32\textwidth]{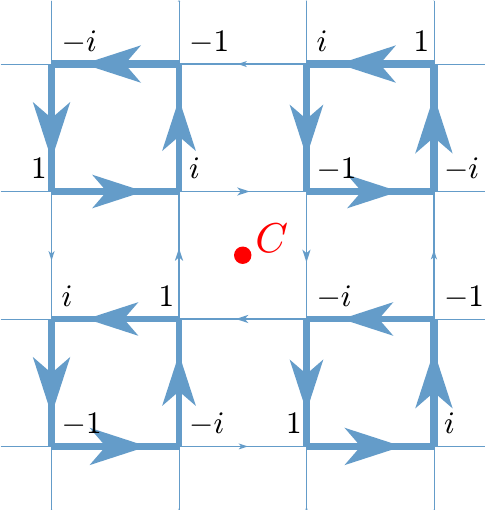}% Here is how to import EPS art
\caption{\label{fig:stringphase} The Bloch-wave phase factor on each site of the $L_z = 1$ ground state. The local angular momentum of the $2\times 2$ closed-string structure is $L_z(\text{string}) = -1$. The nearby two closed strings have $\pi$ phase shift, which is consistent with the global $L_z = 1$ symmetry under rotation around the system center $C$. }
\end{figure}

In this appendix, we use ED to investigate how the ground state quantum numbers evolve with $t/J$ in a minimal $2\times 2$ single-hole-doped system with total $S^z = -1/2$.

ED calculations reveal that in the small $t$ limit ($t/J<0.5$), the ground state is unique and carries angular momentum $L_z = 2$. It can be expressed as: 
\begin{equation}
|\Psi_{\mathrm{st}}\rangle_{1\mathrm{h}} = \sum^4_{i=1} \frac{1}{\sqrt{2}} c_{i\uparrow} |\phi_0\rangle, \label{eq:EDtsmall}
\end{equation}
where $|\phi_0\rangle$ is the half-filling ground state of the $2\times 2$ system (which itself has $L_z = 2$ due to fermion statistics, as shown in Fig.~\ref{fig:EDsolution}(a)) and $1/\sqrt{2}$ is a normalization factor. In this limit where $J \gg t$, minimizing superexchange energy is prioritized. When the hole is fixed at a given site, $c_{i\uparrow}|\phi_0\rangle$ represents sudden removal of a $\uparrow$-spin without additional spin distortion, thus preserving antiferromagnetic energy. For hole hopping, the kinetic energy is minimized when the quasiparticle amplitudes $c_{i\uparrow}|\phi_0\rangle$ at different sites are in phase, leading to $E_t = -t$, consistent with the form in Eq.~(\ref{eq:EDtsmall}).

For intermediate $t/J$ ratios ($0.5 < t/J < 3.8$), the ground states become doubly degenerate with $L_z = \pm 1$ (equivalent to momenta $k = \pm \pi/2$ in a 1D picture). Following Eq.~(\ref{eq:realpwave}), we combine these into real-valued wave functions; the $|p_{yx}\rangle$ state is shown in Fig.~\ref{fig:EDsolution}(b).

The ground state exhibits two distinct components: a quasiparticle component on sites 2 and 4 ($c_{2\uparrow}|\phi_0\rangle$ and $c_{4\uparrow}|\phi_0\rangle$), and an incoherent component involving spin twists $\hat{\Pi}_i$ on sites 1 and 3 ($c_{1\uparrow}\hat{\Pi}_1|\phi_0\rangle$ and $c_{3\uparrow}\hat{\Pi}_3|\phi_0\rangle$).

Here, the operator $\hat{\Pi}_i$ represents a specific spin excitation or ``twist" on $|\phi_0\rangle$. Physically, $\hat{\Pi}_i$ induces a diagonal spin-singlet pairing between the two sites on the sublattice opposite to the hole (e.g., if the hole is at site 1 on the A-sublattice, $\hat{\Pi}_1$ forms a singlet between sites 2 and 4 on the B-sublattice). This contrasts with the nearest-neighbor singlet structure of $|\phi_0\rangle$. Consequently, one can verify that $\langle \phi_0 |c^{\dagger}_{i\uparrow} c_{i\uparrow}\hat{\Pi}_i |\phi_0\rangle = 0$, ensuring orthogonality between the incoherent and quasiparticle components even when the hole resides on the same site.

The presence of this incoherent component increases the overlap with $\hat{H}_t c_{i\uparrow}|\phi_0\rangle$ (the hopping term acting on the quasiparticle part), enhancing kinetic energy gain through resonance between the two components—a key aspect of hole dynamics. The sign change under $\pi$ rotation of this $|p_{yx}\rangle$ state, corresponding to $L_z = \pm 1$ in the $C_4$-symmetric state, stems from the phase-string sign structure: In the antiferromagnetic environment, a two-step hole hopping process typically exchanges the hole with two opposite spins ($S^z$), accumulating a phase $(-1)=(\pm 1)\times(\mp 1)$ in the wave function.

Figure~\ref{fig:EDsolution}(c) shows the expression of the basis state $c_{i\uparrow}e^{-im(\hat\Omega_i-\hat\Omega_v)}|\phi_0\rangle$ ($i=2$) used in our variational wave function Ansatz [Eq.~(\ref{eq:singlehole})], with $v$ at the plaquette center. Both the quasiparticle component $c_{i\uparrow}|\phi_0\rangle$ and the incoherent component $c_{i\uparrow}\hat{\Pi}_i |\phi_0\rangle$ are recovered in this basis. The different interference patterns between the $m = \pm 1$ bases (e.g., $\sum_{m=\pm 1} m c_{i\uparrow}e^{-im(\hat\Omega_i-\hat\Omega_v)}|\phi_0\rangle$ yields only the incoherent component) explain the $v$-dependent behavior of the 2D $|p_{xy}(v)\rangle$ state noted in the main text (cf. Fig.~\ref{fig:pwave}).

In the large $t$ limit ($t>3.8$), ED yields a unique $L_z = 0$ ground state:
\begin{equation}
\begin{split}
|\Psi_{\mathrm{lt}}\rangle_{\text{1h}} = \frac{1}{\sqrt{12}} \sum^4_{i=1} \big( & c^{\dagger}_{i+1\uparrow} c^{\dagger}_{i+2\downarrow} c^{\dagger}_{i+3\downarrow} + c^{\dagger}_{i+1\downarrow} c^{\dagger}_{i+2\uparrow} c^{\dagger}_{i+3\downarrow} \\
& + c^{\dagger}_{i+1\downarrow} c^{\dagger}_{i+2\downarrow} c^{\dagger}_{i+3\uparrow} \big) |0\rangle,
\end{split}
\label{eq:EDtlarge}
\end{equation}
where $ c_{i+4,\sigma} \equiv c_{i,\sigma}$ and the $|0\rangle$ denotes the vacuum state. This equal-weight superposition over Ising basis states minimizes the kinetic energy to $E_t = -2t$ (the theoretical minimum), analogous to Nagaoka polaron physics \cite{PhysRev.147.392}.

Finally, we present the ground state of the two-hole-doped system in Fig.~\ref{fig:EDsolution}(d). This state exhibits a pairing structure consistent with our discussion in the main text. Specifically, it comprises a NN Cooper-pair component (first bracket) and an incoherent component (second bracket). The latter features NNN pairing accompanied by a same-sublattice singlet pair, which acts as a spin excitation.

\section{\label{app:phase}The momentum of the ``closed string'' structure}

\begin{figure*}[h]
\includegraphics[width=0.7\textwidth]{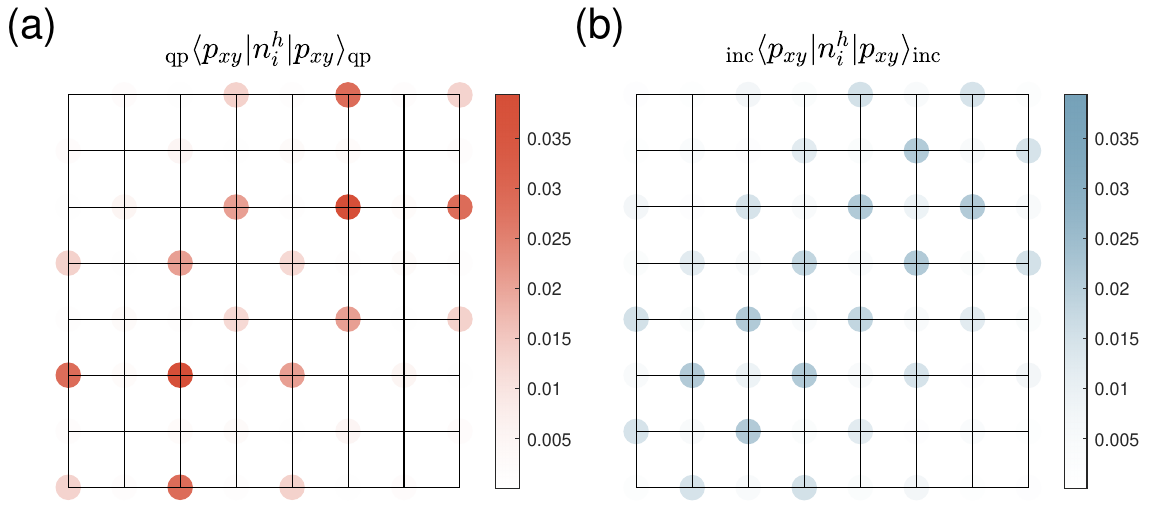}% Here is how to import EPS art
\caption{\label{fig:qp_incoh_lx8} The hole densities of the (a) quasiparticle component and (b) incoherent component of the $|p_{xy}\rangle$ real function.}
\end{figure*}

\begin{figure*}[h]
\includegraphics[width=0.7\textwidth]{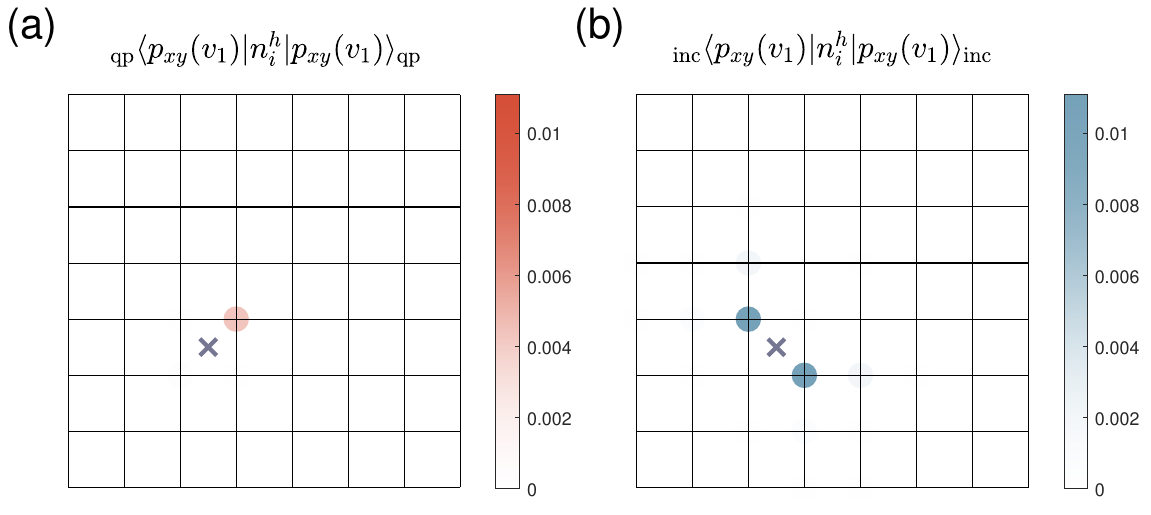}% Here is how to import EPS art
\caption{\label{fig:incoh_dominate_lx8} The hole densities of the (a) quasiparticle component and (b) incoherent component of the $|p_{xy}(v_1)\rangle$ real function. The projected antimeron position $v_1$ is labeled by the black cross.}
\end{figure*}

\begin{figure*}[h]
\includegraphics[width=0.7\textwidth]{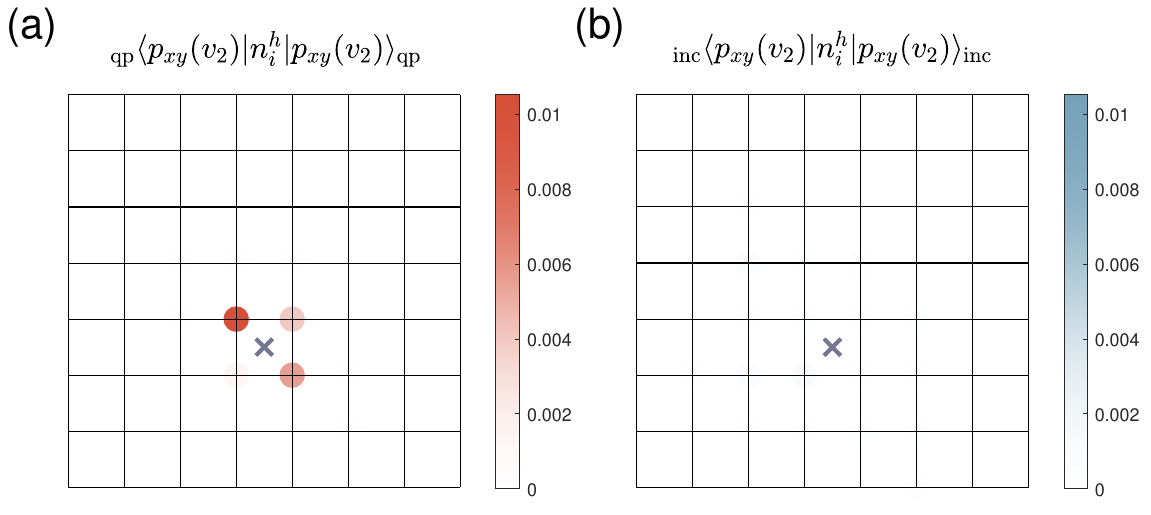}% Here is how to import EPS art
\caption{\label{fig:qp_dominate_lx8} The hole densities of the (a) quasiparticle component and (b) incoherent component of the $|p_{xy}(v_2)\rangle$ real function. The projected antimeron position $v_2$ is labeled by the black cross.}
\end{figure*}

In Section \ref{sec:3}, we identified a $2 \times 2$ closed-string-like excitation in the single-hole ground state. The emergent degree of freedom, characterized by the local quantum number $L_z = \pm 1$ of the closed-string excitation, remains robust when the system size significantly exceeds the string size, as boundary effects diminish. Based on this, we investigate the tunneling behavior of the closed string below.

To elucidate the relative phase between neighboring closed strings, we plot the argument of the quasiparticle wave function component $\text{Arg}(\varphi(i))$ for $|\Psi_{\text{qp}}\rangle_{1\mathrm{h}} = \sum_i\varphi(i)c_{i\uparrow} |\phi_0\rangle$ on an $8\times 8$ lattice, shown in Fig.~\ref{fig:stringphase}. We find that the local angular momentum of the closed strings is $L_z(\text{string}) = -1$. This contrasts with the $L_z = +1$ of the full wave function. This difference arises because the closed string carries momentum $(\frac{\pi}{2},\frac{\pi}{2})$ in the folded Brillouin zone, rather than $(0,0)$. The $(\frac{\pi}{2},\frac{\pi}{2})$ momentum implies a $\pi$ phase shift when the closed string translates by $2a_0$ along $\hat{x}$ or $\hat{y}$ direction. 
Accounting for this $\pi$ phase shift implies that the closed-string pattern exhibits a minimal spatial periodicity of $4a_0 \times 4a_0$.

In the $6\times 6$ system, the angular momentum of the closed string coincides with that of the full wave function, as the string resides at the system center (see Fig.~\ref{fig:currentlx6}). This explains the differing local chiralities observed in the hole current patterns between $L_z = 1$ ground states on the $6\times 6$ and $8\times 8$ lattices.

The phase pattern shown in Fig.~\ref{fig:stringphase} precisely matches the phase $\phi_i$ associated with the following superposition of Bloch-wave states (up to an arbitrary U(1) phase factor):
\begin{equation}
\frac{1}{2}\left(c_{K_1\uparrow}-ic_{K_2\uparrow}-c_{K_3\uparrow}+ic_{K_4\uparrow}\right) |\phi_0\rangle \equiv \frac{1}{\sqrt{N}} \sum_{i} \phi_i c_{i\uparrow}|\phi_0\rangle \label{eq:Lz1quasi}
\end{equation}
where $K_1 = (\frac{\pi}{2},\frac{\pi}{2})$, $K_2 = (-\frac{\pi}{2},\frac{\pi}{2})$, $K_3 = (-\frac{\pi}{2},-\frac{\pi}{2})$, $K_4 = (\frac{\pi}{2},-\frac{\pi}{2})$. Therefore, the quasiparticle component at momentum $(\pm\frac{\pi}{2},\pm\frac{\pi}{2})$ can be viewed as emerging from the $(\frac{\pi}{2},\frac{\pi}{2})$ momentum inherent to the closed string excitation.

\section{\label{app:pwave}The emergent $p$-wave orbital in the single-hole incoherent component}

In this appendix, we show the density distributions of quasiparticle and incoherent components of the real wave function $|p_{xy}\rangle$ in Fig.~\ref{fig:qp_incoh_lx8}, which shows the staggered pattern manifested in Fig.~\ref{fig:pwave}, i.e., the two components mainly distribute on the different sublattices. When we project the antimeron at different positions $v_1$ and $v_2$, the incoherent and quasiparticle component dominates in $|p_{xy}(v_1)\rangle$ and $|p_{xy}(v_2)\rangle$ respectively [cf. Fig.~\ref{fig:incoh_dominate_lx8} and Fig.~\ref{fig:qp_dominate_lx8}], which is consistent with the incoherent component ratio $I(v)$ results shown in Fig.~\ref{fig:antimeron}(a). The incoherent component of $|p_{xy}(v)\rangle$ dominates if $v$ is chosen on the loop-current centers of the complex $L_z = \pm 1$ wave functions.

\clearpage
\bibliography{apssamp}% Produces the bibliography via BibTeX.

\end{document}